\documentclass[scriptaddress,twocolumn,pre,showkeys]{revtex4-1}

	\usepackage{bm}
	\usepackage{amsmath,amssymb} 
	\usepackage{makeidx}
	\usepackage{amsfonts}
	\usepackage[ansinew]{inputenc}
	\usepackage[usenames,dvipsnames]{pstricks}
	\usepackage{subfigure}
	\usepackage{epsfig}
	\usepackage[colorlinks,hyperindex]{hyperref}
	\usepackage{tikz}
	\usepackage[T2A,T1]{fontenc}
\newcommand{\yaf}{\mbox{\usefont{T2A}{\rmdefault}{m}{n}\CYRYA\CYRF}}
	\hypersetup
	{
		colorlinks,%
		citecolor=black,%
		linkcolor=black,%
		urlcolor=black,%
	}

\newcommand{\beq}{\begin{equation}}
\newcommand{\eeq}{\end{equation}}

\makeindex

\begin{document}
\title{Magnetization of laser-produced plasma in a chiral hollow target}
\author{Ph.Korneev}
\email{korneev@theor.mephi.ru}
\address{NRNU MEPhI, Moscow 115409, Russian Federation}
\author{V. Tikhonchuk, E. d'Humi\`eres}
\address{University of Bordeaux, CNRS, CEA, CELIA, 33405 Talence, France}

\begin{abstract}
It is demonstrated that targets with a broken rotational symmetry may facilitate generation of a strong axial (poloidal) magnetic field. An intense laser beam irradiating such a target creates intense electron currents carrying vorticity and producing strong spontaneous magnetic fields. Combined with a laser electron acceleration, such targets may be used for generation and guiding of magnetized, collimated particle or plasma beams. 
\end{abstract}

\maketitle
\section{Introduction}\label{sec1}

Laser acceleration of charged particles is a research domain promising interesting applications in the fundamental science, medicine and technology \cite{esarey-revmodphys09, macchi-revmodphys13, pukhov-nature2004}. Both electrons and ions may be accelerated with lasers, but so far produced beams unfortunately have a broad energy and angular distribution much larger than in conventional charged particle accelerators. This is explained by a poorly controlled injection of particles in the acceleration phase, a complicated structure of laser induced accelerating fields and a high beam charge. This is especially true for the electrons because of their small charge-to-mass ratio. 

Magnetic fields are widely used for a collimation and guiding of particle beams, but the common methods of generation of strong magnetic fields with pulse power systems are hardly compatible with laser particle accelerators because of significantly different spatial and temporal scales \cite{albertazzi-science14, santos-njp15}. In contrast, the laser generated magnetic fields could be much better suited for the beam manipulation, but the existing methods of creation of laser-generated magnetic fields are not sufficiently developed and the spatial structure of such fields is usually limited to the poloidal magnetic field component \cite{volpe-pre14, cai-pre11, robinson-pop07, ramakrishna-prl10}. It is desirable to develop more advanced methods of controllable magnetic field generation with lasers, which would be compatible with the laser generated accelerating fields. 

A use of chiral structures allows to affect the polarization state of laser beams. It was suggested in Ref. \cite{shi-prl14} that laser beams reflected from a helically-shaped target may acquire an orbital angular momentum. The recoil orbital momentum is transferred to the target electrons and dissipated. A similar setup was considered in Ref. \cite{lecz-lpb15}. More generally, by an appropriate choice of the geometry of interaction, target shape and conductivity, one may extend the life time of electron currents and use them for generation of a quasi-static magnetic field. One example of such a structure was suggested in Ref. \cite{korneev-pre15} where a laser beam is injected obliquely in a snail-shaped structure thus driving a vortical electron current in the laser propagation direction associated with a strong magnetic field oriented perpendicularly to the laser propagation axis.  Here we propose another interaction geometry where the magnetic field is generated in the direction of laser beam propagation. It allows to significantly increase the interaction length and to combine together the processes of laser particle acceleration and magnetic field generation. 

\begin{figure}{\begin{center}\includegraphics[width=0.34\textwidth]
{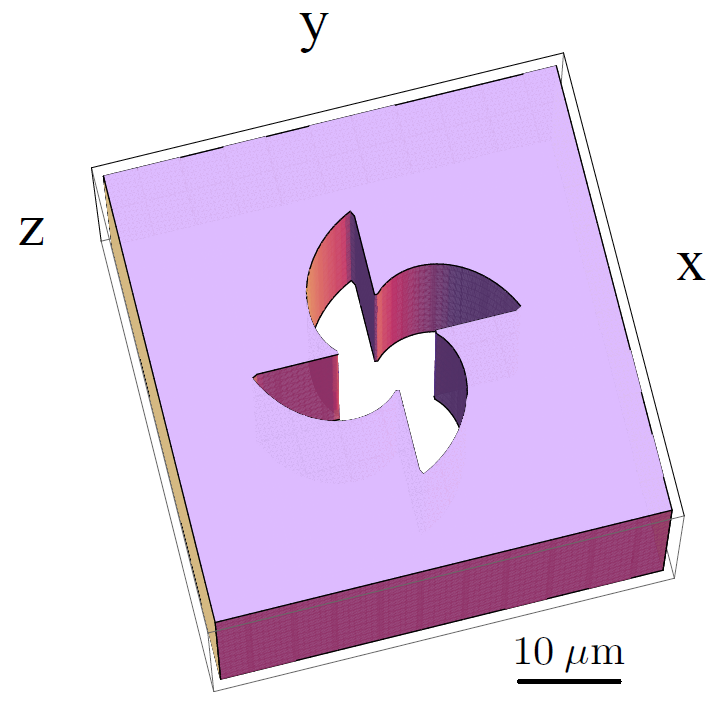}\end{center}}
{\begin{center}\includegraphics[width=0.34\textwidth]
{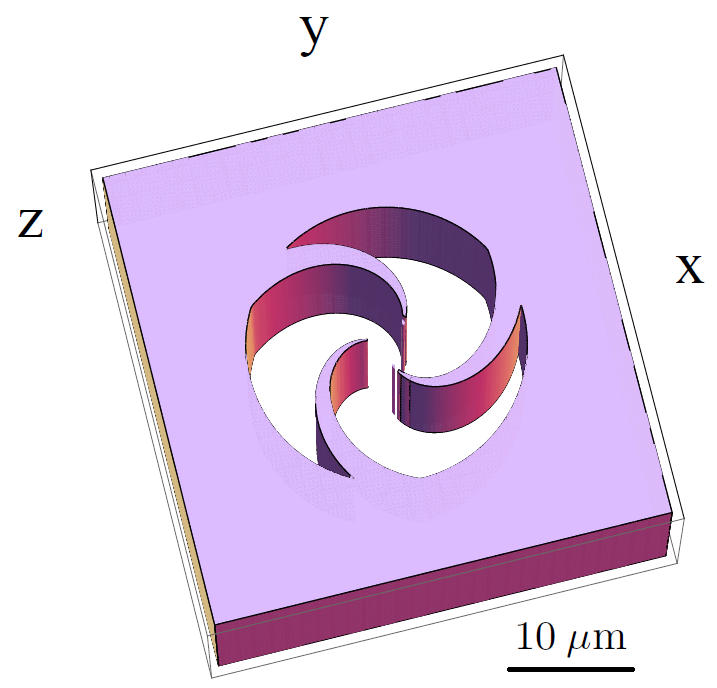}\end{center}}
\caption{Examples of targets with a broken rotational symmetry considered in the paper. {The characteristic size of the hole is 10 $\mu$m.} Upper panel refers to the target T1, bottom panel -- to the target T2. According to the definition of chirality, Eq.(\ref{def_eta_tgt}), it is positive for T1 and negative for T2.}
\label{target}
\end{figure}

Two examples of hollow targets with a broken rotational symmetry are shown in Fig. \ref{target}. In contrast to toroidal magnetic fields generated by a straight electron current, here an azimuthal current having a structure resembling a charged $\theta$-pinch generates an axial magnetic field inside the target hole. Although the same laser pulse may do both, the magnetic field generation and particle acceleration, we focus here on formation of the magnetized structure leaving the acceleration physics for a separate publication. 

The paper is organized as follows: first, we present in Sec. \ref{sec2A} a simplifying approach allowing to reduce the numerical problem from the fully three-dimensional geometry to a planar geometry with three velocity components (2D3V). The main elements of our scheme of the axial magnetic field generation are illustrated in Sec. \ref{sec2B} with the Particle-In-Cell (PIC) simulations of a laser pulse interaction with a hollow chiral target. By using the targets shown in Fig. \ref{target} we analyze typical current and magnetic field structures and their temporal evolution. Then in Sec. \ref{sec3} we discuss the properties of the vortical structure and the potential applications to the control of particle acceleration and guiding.

\section{Particle-In-Cell simulations of a laser pulse interaction with a hollow target}\label{sec2}
The problem of laser pulse interaction with an azimuthally asymmetric target is fully three dimensional and thus presents serious difficulties for both the theoretical analysis and numerical simulations. It can be simplified by reduction to a two-dimensional planar geometry considering one slice in the plane perpendicular to the laser propagation direction $z$. That means that the plasma parameters and laser field variation lengths in the axial direction are much larger that the target hole size and there is no backscattered wave. Moreover, the laser pulse absorption length is supposed to be large compared to the hole size and thus processes in consecutive slices are similar.  Below we present the results of numerical PIC simulations performed with the code PICLS \cite{Sentoku-jcp08} in 2D3V geometry.  

\subsection{Reduction of a 3D problem: 2D slicing of 3D target}\label{sec2A}

Let us consider an electromagnetic wave propagating along $z-$axis with a frequency  $\omega_0$ and a wave vector $k_0=\omega_0/c$:
\beq\mathbf E (z,t)=\mathbf E_0 \exp[i k_0 z-i\omega_0 t],  \label{E(z,t)}\eeq
where $\mathbf E_0$ is the laser electric field amplitude in the $(x,y)$  plane and $c$ is the light velocity.  
\begin{figure}
  \begin{tikzpicture}[x=0.51\linewidth,y=0.45\linewidth]
    \path
    (0,0) node{\includegraphics[width=0.5\linewidth]{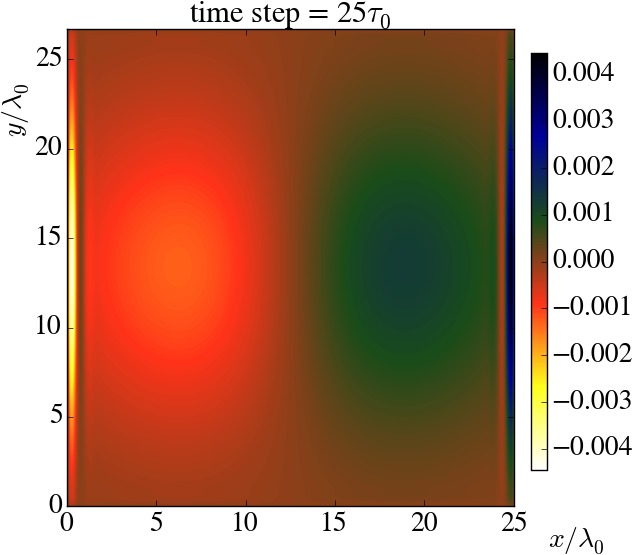}}
    (0,1) node{\includegraphics[width=0.5\linewidth]{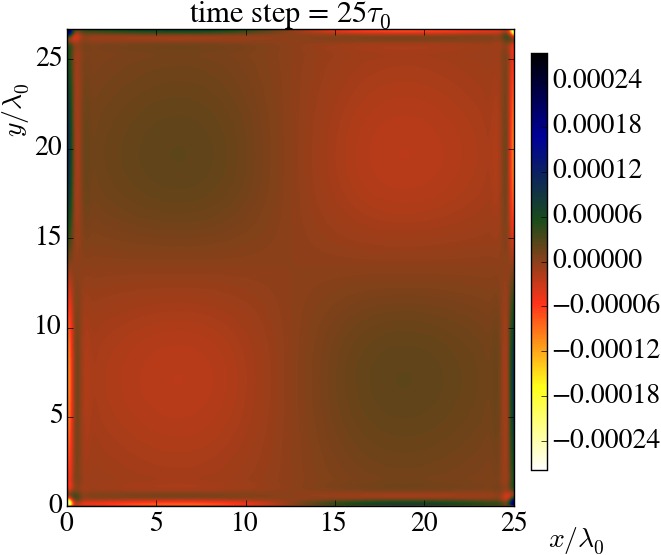}}
    (0,2) node{\includegraphics[width=0.5\linewidth]{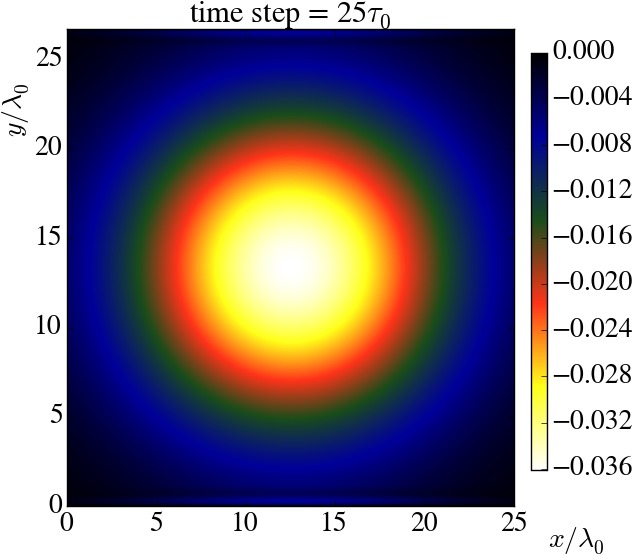}}
    (0.1,0.3) node { \textcolor{white}{\large{$E_z$}}}
    (0.1,1.3) node { \large{$E_y$}}
     (0.1,2.3) node {\textcolor{white}{\large{ $E_x$}}}

    (1,0) node{\includegraphics[width=0.5\linewidth]{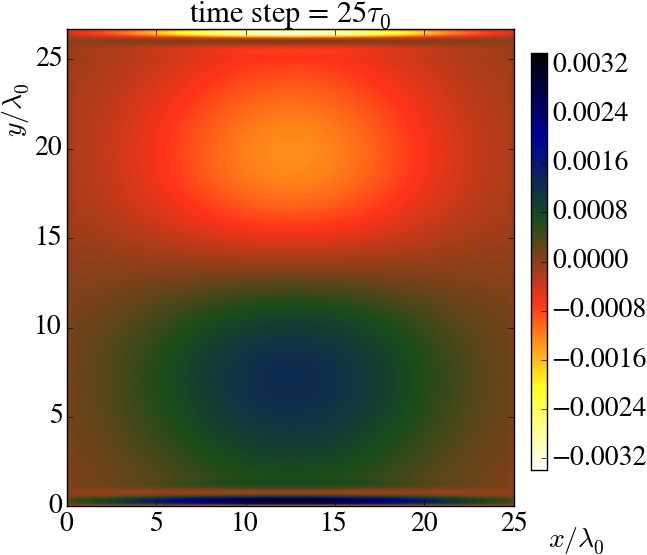}}
    (1,1) node{\includegraphics[width=0.5\linewidth]{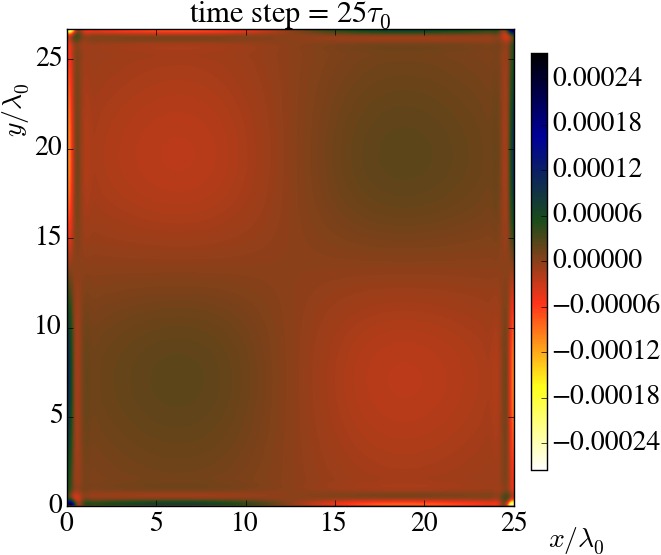}}
    (1,2) node{\includegraphics[width=0.5\linewidth]{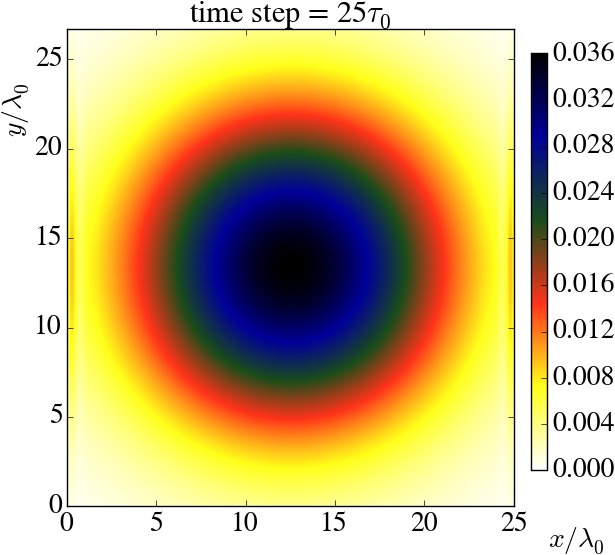}}
    (1.1,0.3) node { {\large{$B_z$}}}
    (1.1,1.3) node {\textcolor{white}{\large{ $B_x$}}}
    (1.1,2.3) node {\textcolor{black}{ \large{$B_y$}}}
;
  \end{tikzpicture}
\caption{Electric and magnetic field components of a small amplitude laser beam propagating in the direction ``into the picture'' in a plasma-free 2D simulation box of a size of 25x25 laser wavelengths. The beam has a width of 20 wavelengths at half maximum, it is linearly polarized and has $E_x$ and $B_y$  components in a free space. Other field components are generated according to the Maxwell's equations due to the boundary conditions in the $x,y$ plane: $E_z$ and $B_z$ components are of a dipole type,  $E_y$ and $B_x$ components are of a quadrupole type, both have much smaller amplitudes than $E_x$ and $B_y$.}
\label{target-free}
\end{figure}

According to the Maxwell's equations, the electric $\mathbf E$ and magnetic $\mathbf B$ fields of a plane wave polarized along $x$-axis are related as $B_y=E_x$. For a laser beam of a limited aperture $\Delta_r\gg 1/k_0$ and a finite duration $\Delta_t\gg1/\omega_0$, these fields present a slow dependence in the perpendicular plane $x,y$ and on time defining the beam spatial and temporal profile. Moreover, other field components are also generated but their amplitudes are smaller by a factor $\epsilon =(k_0\Delta_r)^{-1}\ll 1$. In the numerical examples presented below we consider a laser beam with the transverse width of the order of ten wavelengths. In this case the plane wave approximation is sufficient and other field components are at least $1/\epsilon$ times smaller. This is demonstrated with field maps in a plasma-free simulation in Fig. \ref{target-free} for a small amplitude laser pulse. Two main components, $E_x$ and $B_y$ dominate with the amplitudes $\sim0.04$ in the relativistic units $m_e c \omega_0/e$. The amplitudes of second-order dipole corrections $E_z$ and $B_z$ are an order of magnitude smaller than the main components. Next corrections of the quadrupole type, $E_y$ and $B_x$, are an order of magnitude smaller than the dipole components. 

While propagating inside the hole, the laser beam edges interact with the target walls and create the secondary (scattered) waves. Assuming that the laser field is not too much perturbed, we describe these secondary electromagnetic fields with the 2D Maxwell's equations in the simulation plane and thus neglecting their variation in the laser propagation direction. This approach is certainly rather simplified as the propagation along $z$-axis is not accounted for. Therefore, these fields do not propagate along $z$-axis and remain in the simulation box even after the laser pulse ends. Consequently, the scattered laser wave is not described. However, during the time of laser pulse, the amplitude of secondary fields is evaluated correctly and the interaction of these fields with the plasma is accounted for. They induce charge separation electrostatic fields at the target edges and electron currents having a vortical component.  As the scope of the present paper is limited to description of plasma heating and electron current generation during the laser pulse, such a planar geometry approximation is sufficient for estimates of the magnitude of expected effects.

\subsection{Laser interaction with a hollow target }\label{sec2B}

In numerical simulations we consider a fourth order super-Gaussian pulse profile in time and a Gaussian profile in space
\beq  {E_x}(\mathbf{r},t)= E_0(z,t)\,\exp\left[-\frac{(\mathbf{r}-\mathbf{r_0})^2}{\Delta_r^2}-\frac{(t- t_0)^4}{\Delta_t^4}\right],
\label{heating_laser}\eeq
where $\mathrm{r}_0$ is the center of the target, $t_0$ is a time delay of the pulse, $\Delta_r$ and $\Delta_t$ are the beam width and  the pulse duration.  The laser  wavelength $\lambda_0=0.8~\mu$m corresponds to the period $\tau_0 =2\pi/\omega_0=2.66$ fs and the beam radius $\Delta_r=10\,\lambda_0$. Three laser pulses are considered:  a ``long'' linearly or circularly polarized pulse of 100 periods, $\Delta_t= 100\,\tau_0= 266$ fs with an intensity $I_L=10^{20}$ W/cm$^2$ and a ``short'' linearly polarized pulse $\Delta_t=10\,\tau_0=26.6$ fs with an intensity $I_L=2.5\times10^{21}$W/cm$^2$.  

Two considered targets, shown in Fig. \ref{target} are constituted of a solid cylinder with figure cuts, so that there are four (in T1) or three (in T2) claws, directed toward the center. The claws are disposed periodically and the shape of their edges is defined as
\beq
r( \theta )=r_0\left( 1-\dfrac{\Delta r}{r_0}\dfrac{ \theta }{ \theta _0} \right),
\label{shape}
\eeq
where $r_0,~\Delta r$ and $ \theta _0$ are parameters. For the internal edge (more close to the center) of the T1 target, $r_0= 9~\mu$m, $\Delta r=7.6~\mu$m, and $ \theta _0$ varies from $0$ to $\pi/2$ periodically with a shift of $\pi/2$, while the second edge is straight in the radial direction. Obviously, such a shape brakes the rotational symmetry in the process of interaction. As we prove below, this leads to generation of surface currents of predominant direction, which produce a long-living magnetized plasma structures in the target hollow. For the target T2, the shape of the edges is defined as in Eq.(\ref{shape}), but with the parameters  $r_0=8.4~\mu$m and $\Delta r=5.2~\mu$m for the internal edge, and $r_0=10.9~\mu$m and $\Delta r=10.0~\mu$m for the external edge. In this case, $ \theta _0$ varies from $2\pi/3$ to $0$ periodically with a shift of  $2\pi/3$. In the case of the target T2, we expect currents directed oppositely to those in the target T1, thus the generated internal magnetic field should also have the opposite direction. A quantitative parameter for a target chirality which has opposite signs for the targets T1 and T2 is discussed in Sec. \ref{sec3}.  


Both targets were modeled by homogeneously distributed aluminum ions with a charge $Z=13$ and a  density $n_i=6\times10^{22}$ cm$^{-3}$, and electrons with a density $n_e=7.8\times10^{23}$ cm$^{-3}$ corresponding to a plasma with the density 450 times the critical $n_c=1.5\times10^{21}$ cm$^{-3}$. In all simulations there were 2 ions and 26 electrons per cell, with the initial electron temperature of 50 eV.  The simulation box contained $5748\times6144$ cells, or approximately $25\times27~\mu$m$^2$. The resolution was 230 points per wavelength and per laser period. The collisions were included in the simulations, the collisional module in the code is based on the Takizuka and Abe model \cite{Takizuka-jcp77}.

\subsection{Examples of magnetic structures}

\begin{figure}
  \begin{tikzpicture}[x=0.51\linewidth,y=0.45\linewidth]
    \path
    (0,0) node{\includegraphics[width=0.5\linewidth]{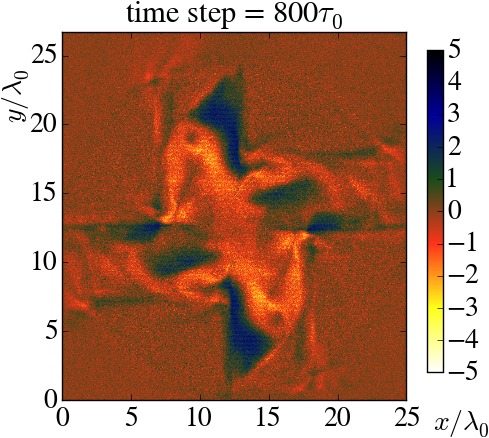}}
    (0,1) node{\includegraphics[width=0.5\linewidth]{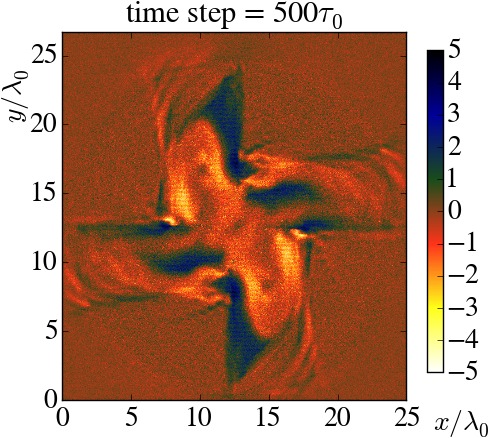}}
    (0,2) node{\includegraphics[width=0.5\linewidth]{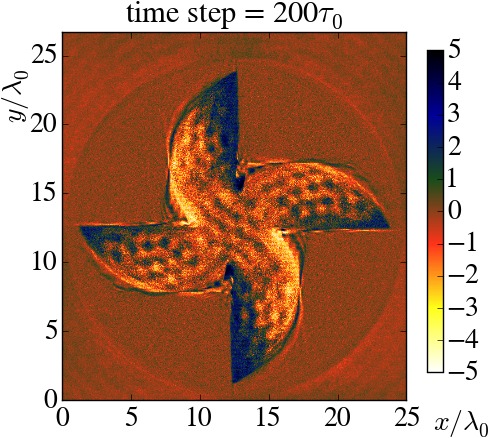}}
    (0,3) node{\includegraphics[width=0.5\linewidth]{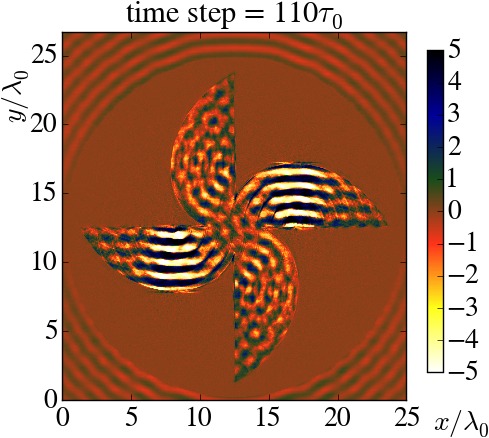}}

    (1,0) node{\includegraphics[width=0.5\linewidth]{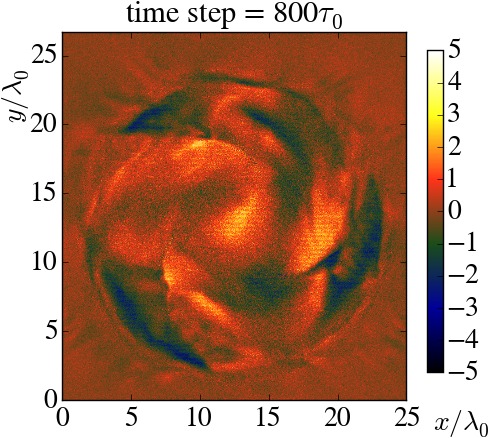}}
    (1,1) node{\includegraphics[width=0.5\linewidth]{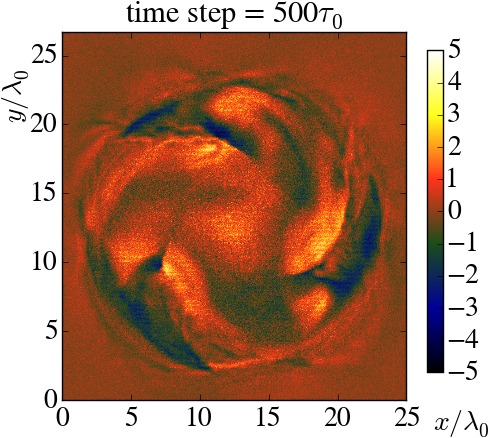}}
    (1,2) node{\includegraphics[width=0.5\linewidth]{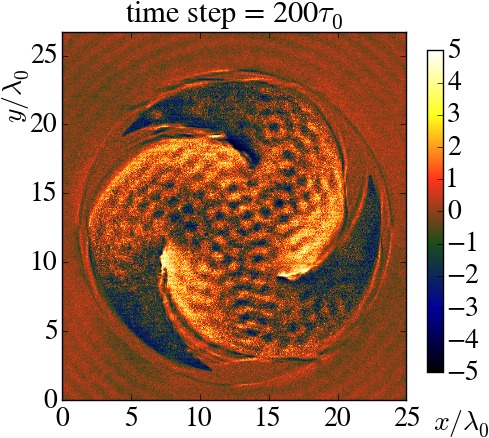}}
    (1,3) node{\includegraphics[width=0.5\linewidth]{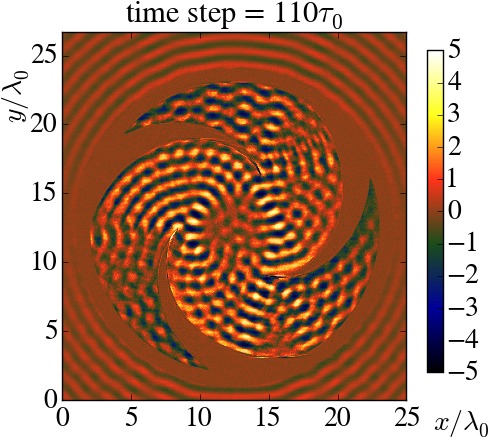}}
;
  \end{tikzpicture}
\caption{Magnetic field $B_z$ in the target T1 (left column) and T2 (right column) at the time moments 110, 200, 500 and $800\,\tau_0$, for a linearly polarized laser pulse with an intensity $I=10^{20}$W/cm$^2$ and duration $\Delta_t=266$ fs. The time $140\tau_0$ corresponds to the laser pulse maximum.}
\label{13-d_14__Bz}
\end{figure}

\begin{figure}
  \begin{tikzpicture}[x=0.51\linewidth,y=0.45\linewidth]
    \path
    (0,0) node{\includegraphics[width=0.5\linewidth]{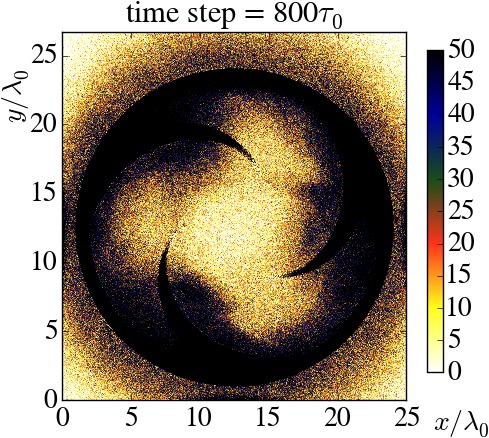}}
    (0,1) node{\includegraphics[width=0.5\linewidth]{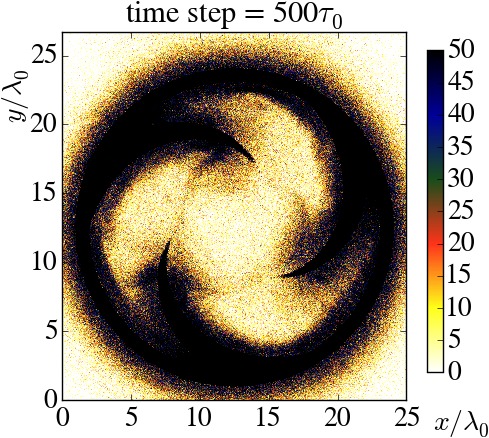}}
    (0,2) node{\includegraphics[width=0.5\linewidth]{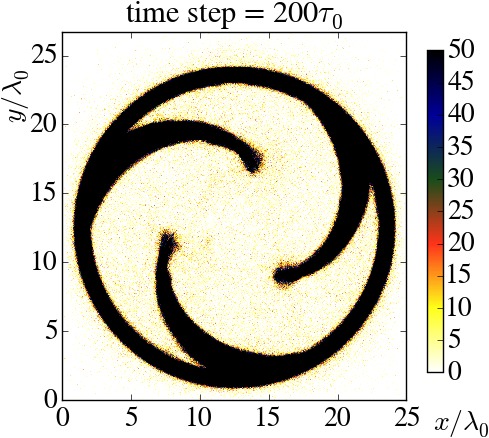}}

    (1,0) node{\includegraphics[width=0.5\linewidth]{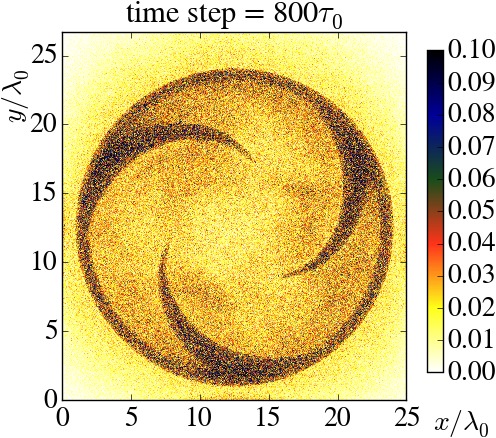}}
    (1,1) node{\includegraphics[width=0.5\linewidth]{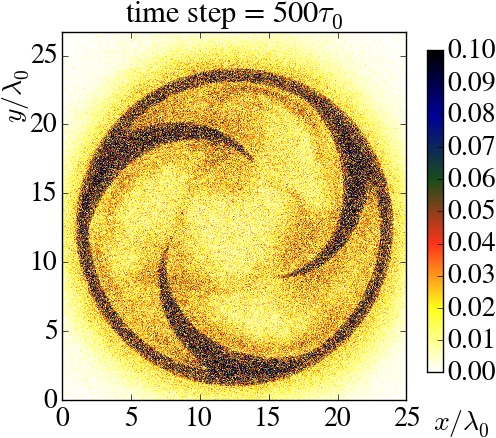}}
    (1,2) node{\includegraphics[width=0.5\linewidth]{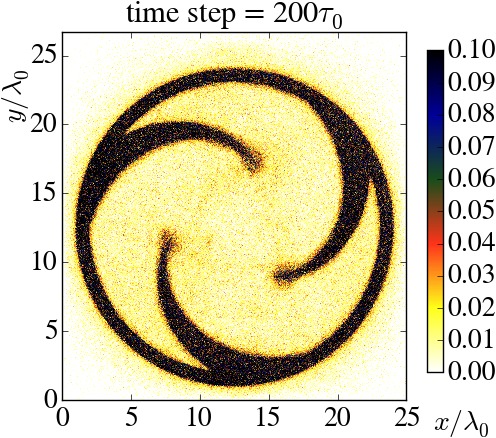}}
;
  \end{tikzpicture}
\caption{Electron density (left column) and absolute value of the current (right column) in the target T2 at time moments 200, 500 and $800\,\tau_0$, for a linearly polarized laser pulse with an intensity $10^{20}$\,W/cm$^2$ and duration 266 fs.}
\label{14__edens_rj}
\end{figure}

\begin{figure}
  \begin{tikzpicture}[x=0.51\linewidth,y=0.45\linewidth]
    \path
    (0,0) node{\includegraphics[width=0.48\linewidth]{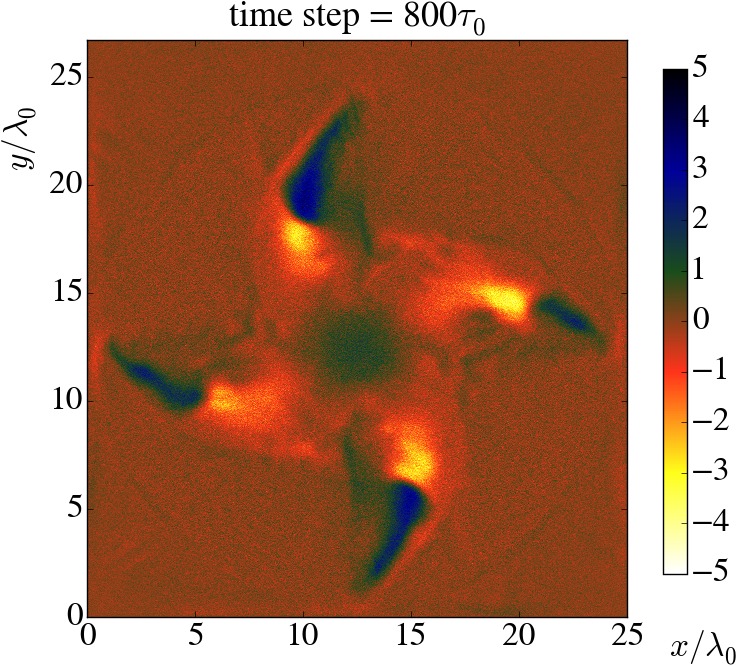}}
    (0,1) node{\includegraphics[width=0.48\linewidth]{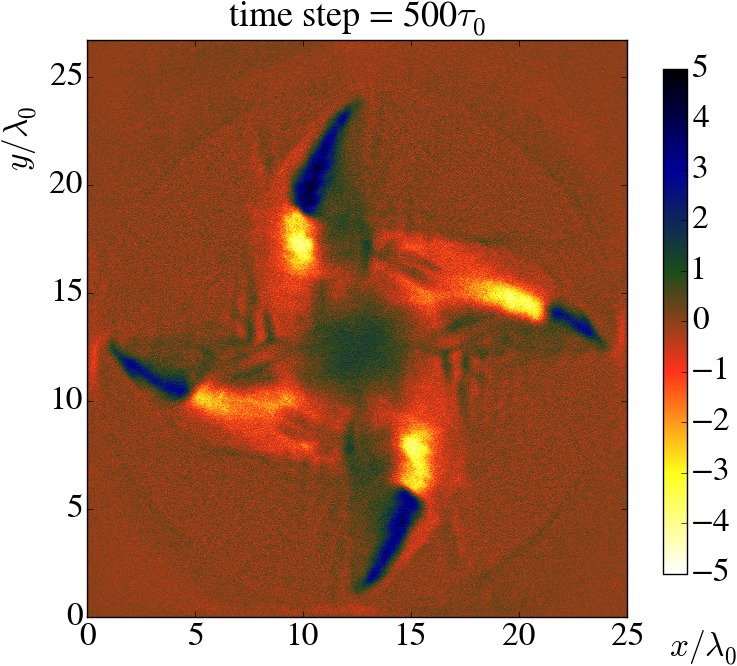}}
    (0,2) node{\includegraphics[width=0.48\linewidth]{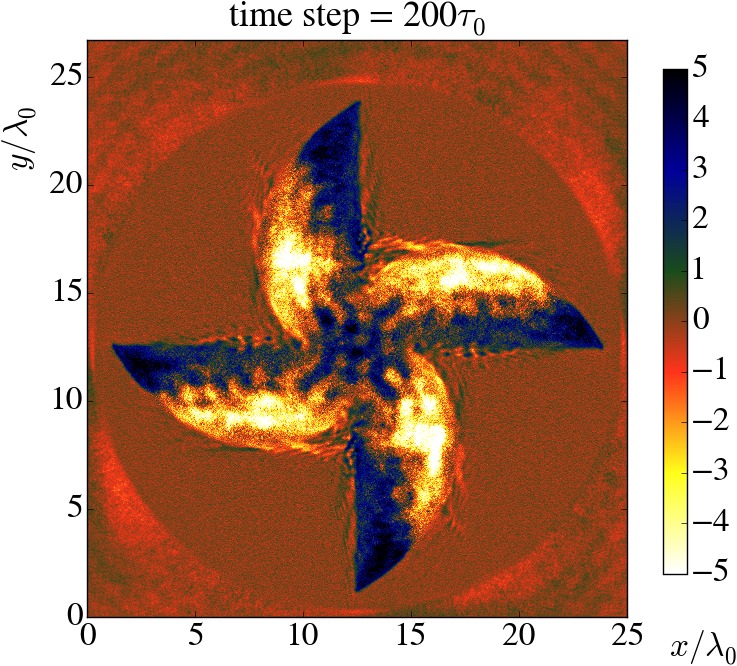}}
    (0,3) node{\includegraphics[width=0.48\linewidth]{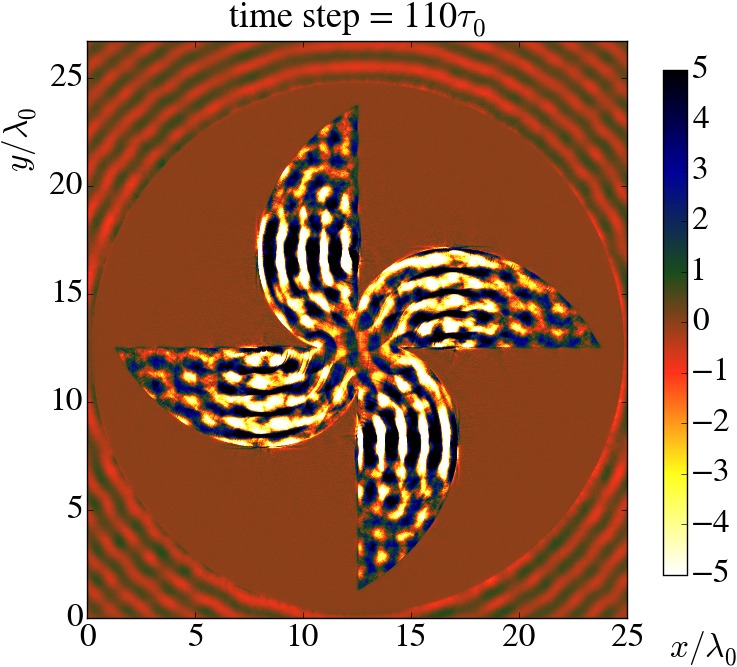}}


    (1,0) node{\includegraphics[width=0.5\linewidth]{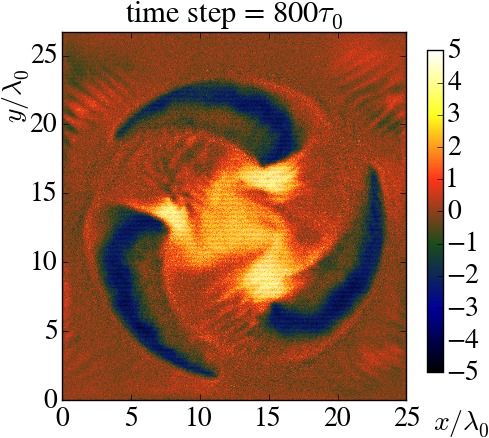}}
    (1,1) node{\includegraphics[width=0.5\linewidth]{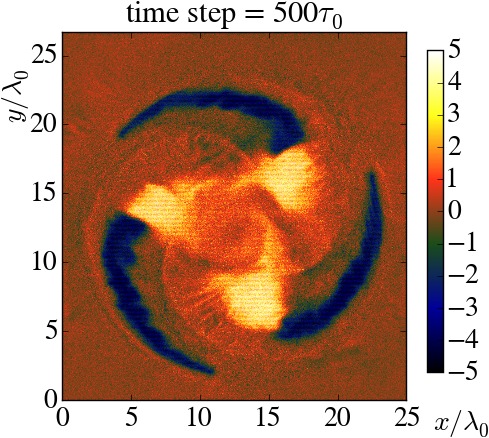}}
    (1,2) node{\includegraphics[width=0.5\linewidth]{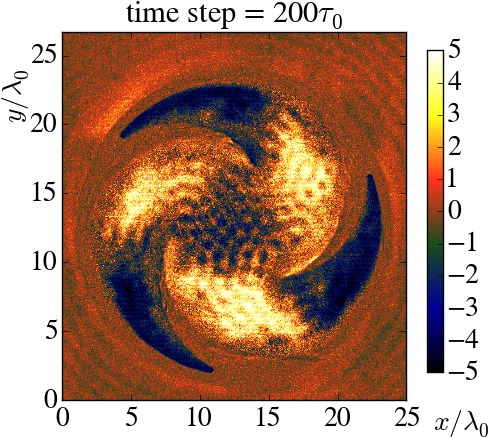}}
    (1,3) node{\includegraphics[width=0.5\linewidth]{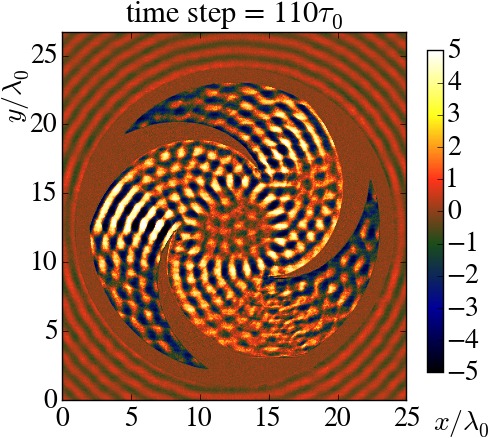}}


;
  \end{tikzpicture}
%
%
\caption{Axial magnetic field for the target T1 (left column) and T2 (right column) at time moments 200, 500 and $800\,\tau_0$, for a circularly polarized pump laser pulse with an intensity $10^{20}$\,W/cm$^2$ and duration 266 fs.}\label{13-f_14-b__Bz}
\end{figure}

Let us consider first a ``long'' linearly polarized pulse. Its interaction with both targets is shown in Fig.\,\ref{13-d_14__Bz}. The laser field structure inside the target hole is defined by the boundary conditions, which are geometrically different for the targets T1 and T2. However, the pulse polarization does not substantially affect the heating process and magnetic field generation. This can be deduced from the comparison of Figs.\,\ref{13-d_14__Bz} and \ref{13-f_14-b__Bz}, which present magnetic fields for both targets for the case of linear and circular polarization respectively. At the beginning of interaction, the electrons heated and ejected from the target walls propagate in the azimuthal direction in the target hole. These electron currents produce  a fast growing axial magnetic field, which survives till the end of simulation, though its amplitude and spatial distribution vary with time. Note, that as expected, the direction of magnetic fields for the targets T1 and T2 is different (be aware of the opposite color scales for T1 and T2). In the hole center for both considered targets at the end of the laser pulse the  magnetic field amplitude is of the order of 1 in the relativistic units. Figure \ref{14__edens_rj} shows the distribution of the electron density and the absolute value of electric current in the target T2 providing an insight on the process of magnetic field generation. The currents are formed firstly near the target internal surface and then move towards the center. Initially, the radial motion dominates, see the time step $200\,\tau_0$ in Fig. \ref{14__edens_rj}. Then, the generated magnetic fields turn electrons in azimuthal direction thus forming a rotating electron flow in the direction defined by the target chirality. Such an annular electron density structure around the central region is shown at the time step $500\,\tau_0$ in Fig. \ref{14__edens_rj}. Later in time, the electron diffusion across the magnetic field makes this structure smoother, as shown in the time step $800\,\tau_0$ in Fig. \ref{14__edens_rj}.

\begin{figure}
  \begin{tikzpicture}[x=0.51\linewidth,y=0.45\linewidth]
    \path
    (0,0) node{\includegraphics[width=0.5\linewidth]{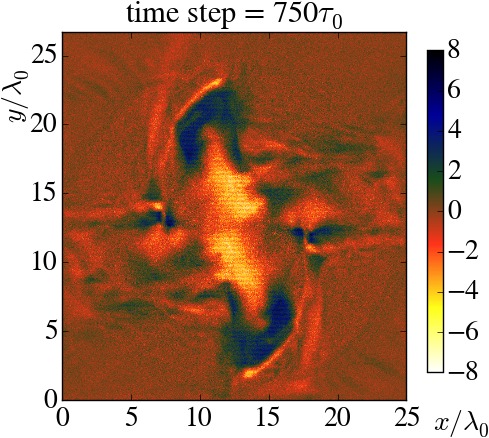}}
    (0,1) node{\includegraphics[width=0.5\linewidth]{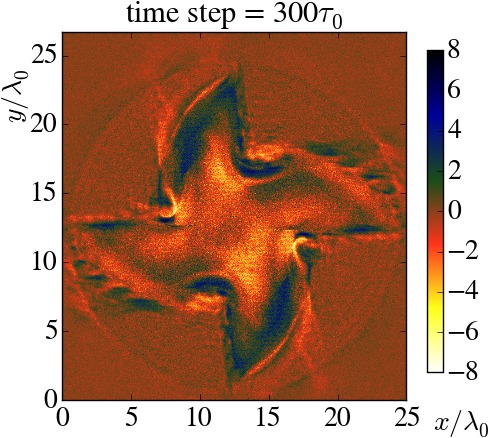}}
    (0,2) node{\includegraphics[width=0.5\linewidth]{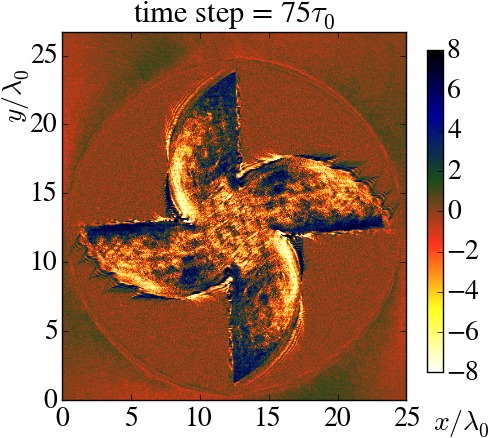}}
    (0,3) node{\includegraphics[width=0.5\linewidth]{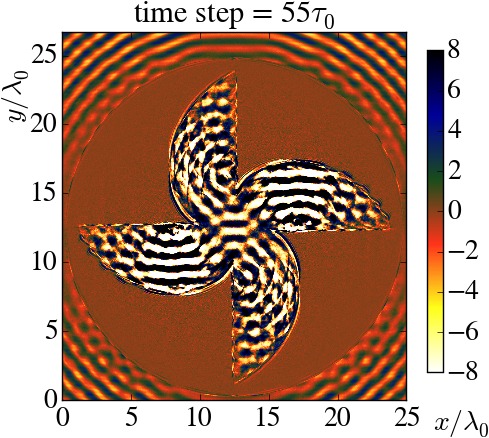}}

    (1,0) node{\includegraphics[width=0.5\linewidth]{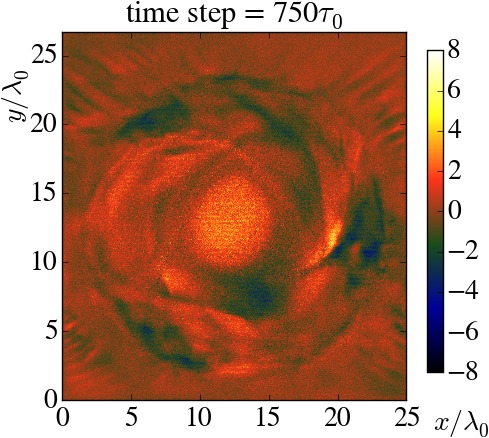}}
    (1,1) node{\includegraphics[width=0.5\linewidth]{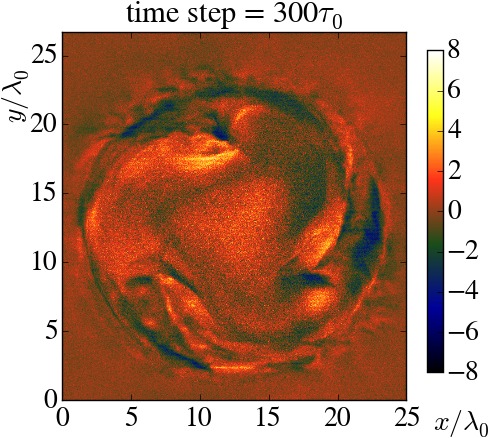}}
    (1,2) node{\includegraphics[width=0.5\linewidth]{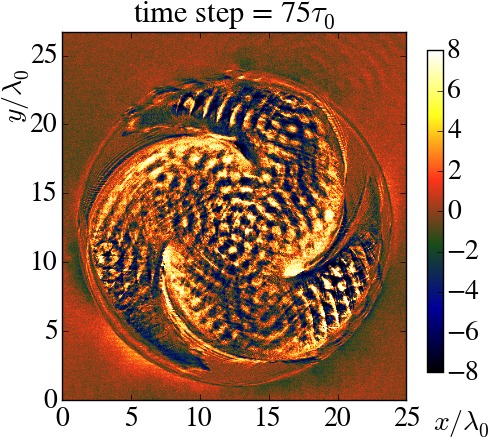}}
    (1,3) node{\includegraphics[width=0.5\linewidth]{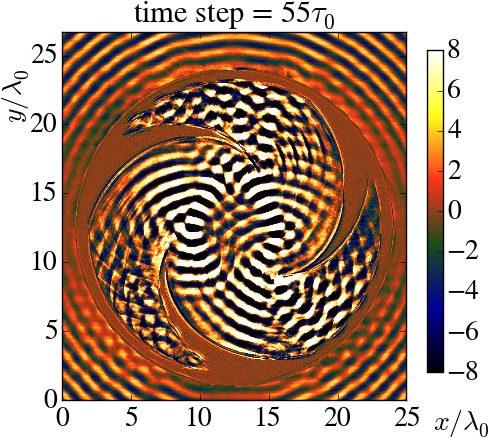}}
;
  \end{tikzpicture}
\caption{Axial magnetic field $B_z$ in the target T1 (left column) and T2 (right column) at time moments 55, 75, 300 and $750\,\tau_0$, for a linearly polarized laser pulse with an intensity $2.5\times10^{21}$\,W/cm$^2$ and duration 27 fs.}
\label{13-e__14-a__Bz}
\end{figure}

\begin{figure}
  \begin{tikzpicture}[x=0.51\linewidth,y=0.45\linewidth]
    \path
    (0,0) node{\includegraphics[width=0.5\linewidth]{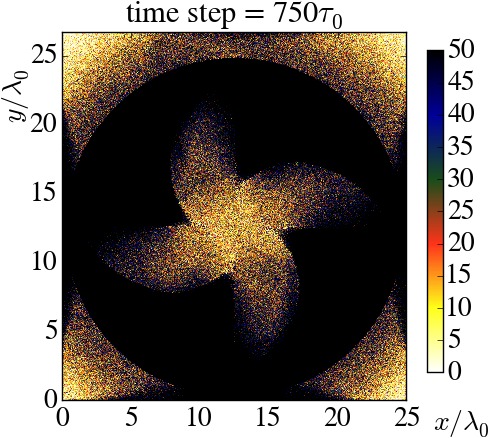}}
    (0,1) node{\includegraphics[width=0.5\linewidth]{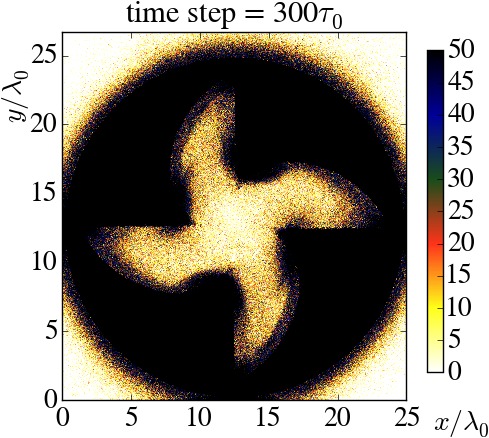}}
    (0,2) node{\includegraphics[width=0.5\linewidth]{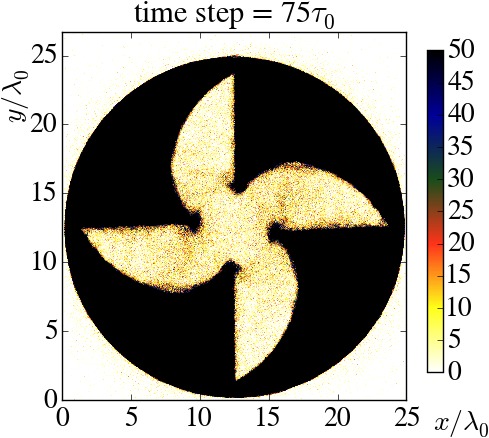}}

    (1,0) node{\includegraphics[width=0.5\linewidth]{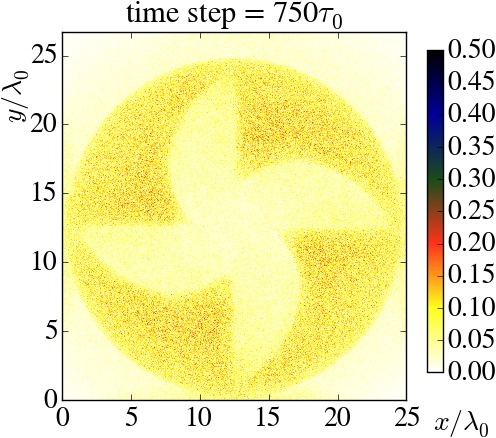}}
    (1,1) node{\includegraphics[width=0.5\linewidth]{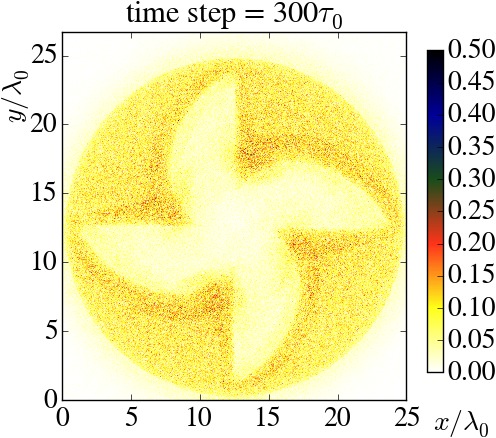}}
    (1,2) node{\includegraphics[width=0.5\linewidth]{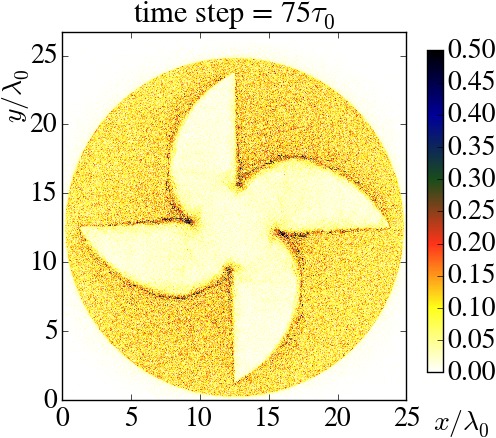}}
;
  \end{tikzpicture}
\caption{Electron density (left column) and absolute value of the current (right column) in the target T2 at  time moments 75, 300 and $750\,\tau_0$, for a linearly polarized pump laser pulse with an intensity $2.5\times10^{21}$\,W/cm$^2$ and duration 27 fs.}
\label{13-e__edens_rj}
\end{figure}

For comparison, Fig. \ref{13-e__14-a__Bz} shows the same targets heated by a linearly polarized ``short'' laser pulse with an intensity $2.5\times10^{21}$ W/cm$^2$ and time duration of 26.6 fs. The processes of electron heating and ejection are rather similar to the previous case, but due to higher electron energies, the resulting magnetic field is more structured and has a higher amplitude. The electron density and current distributions  are shown in Fig. \ref{13-e__edens_rj}  for the target T1  for the same parameters as of Fig. \ref{13-e__14-a__Bz}. The initial radial inward electron motion is seen at the time $75\,\tau_0$, while later the electrons are deviated by the axial magnetic field in the central region. It is worth to mention, that in all the situations, the charge separation fields are neutralized on a time scale of few femtoseconds, so that after the end of laser pulse, the amplitude of electric fields inside the target decreases rapidly.

For the case of a circularly polarized long laser pulse the magnetic field distribution for both targets is shown in Fig. \ref{13-f_14-b__Bz}. A comparison with the previous case for a linearly polarized laser pulse  in Fig. \ref{13-d_14__Bz} shows, that the laser circular polarization enhances the magnetic field amplitude and makes the structure more stable. This can be explained by a more symmetric electron heating and also a more symmetric distribution of plasma flows in the hole. In the presented simulations, we used clock-wise polarization, we did not observe a noticeable difference in the distribution of the magnetic field and electron density by changing polarization direction.

\begin{figure}
{\begin{center}\includegraphics[width=0.5\textwidth]{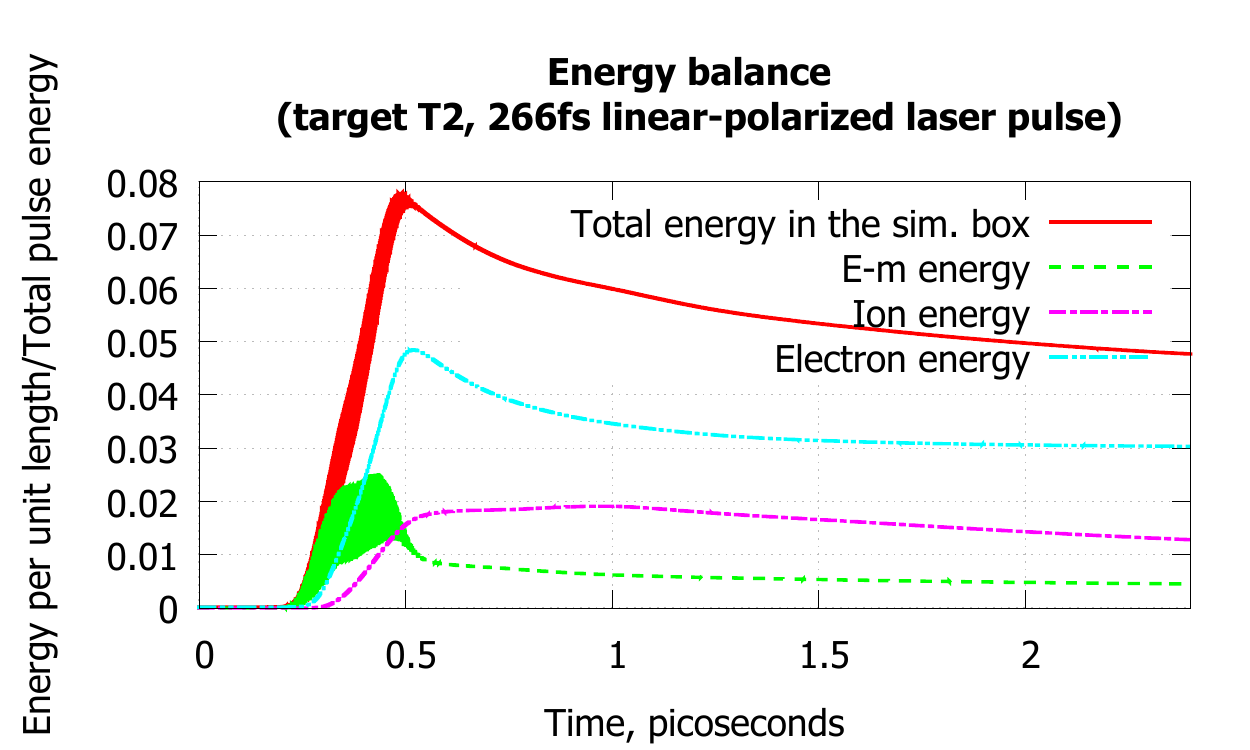}\end{center}} \vspace{-1cm}
{\begin{center}\includegraphics[width=0.5\textwidth]{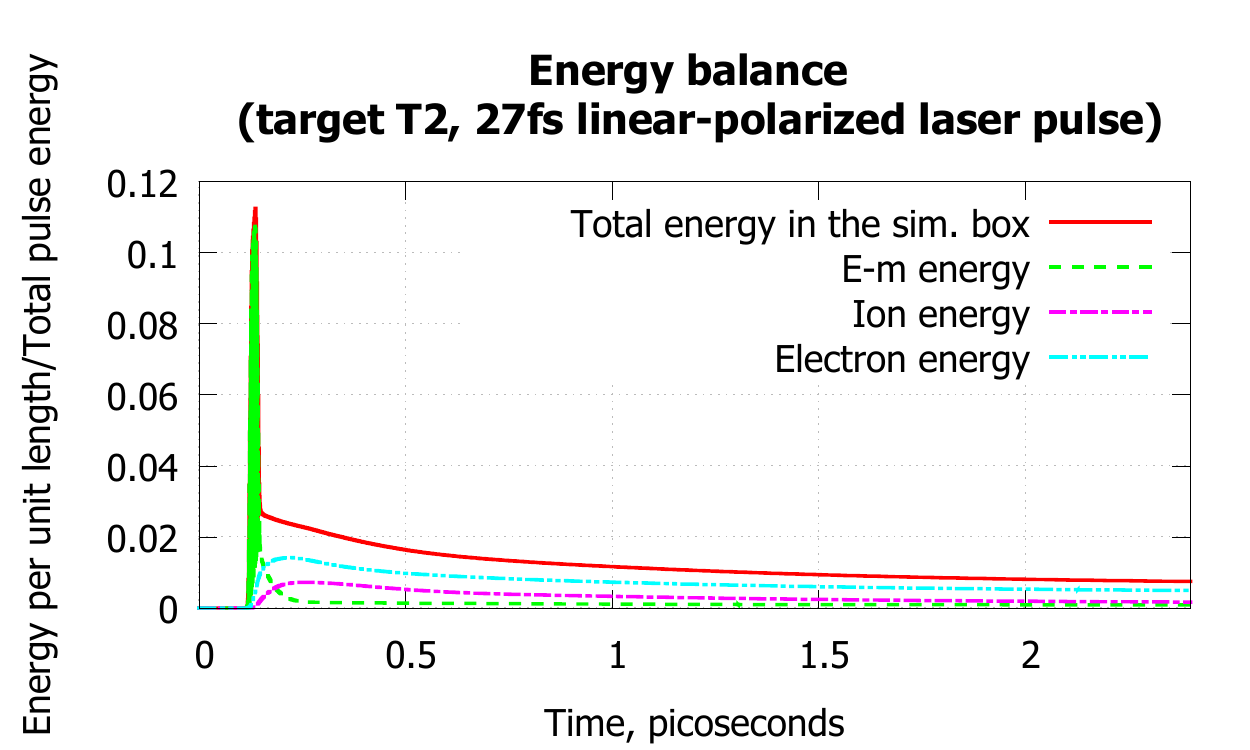}\end{center}}
\vspace{-1cm}
{\begin{center}\includegraphics[width=0.5\textwidth]{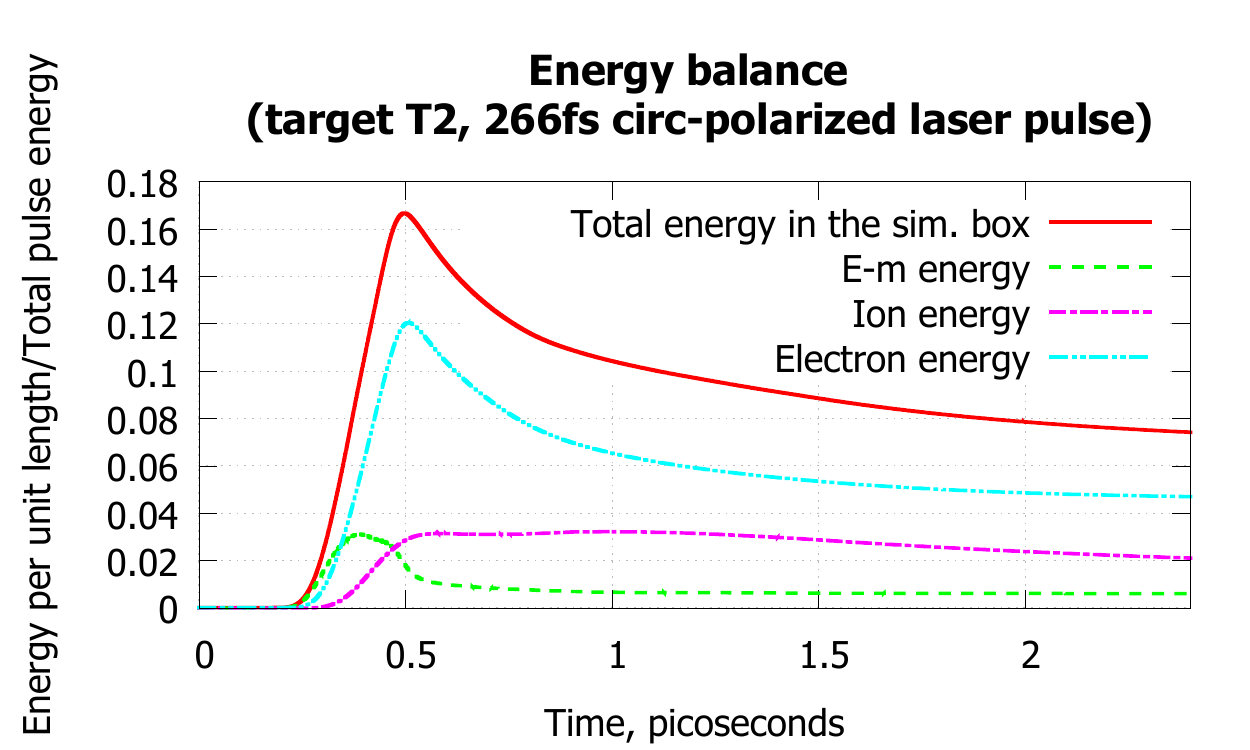}\end{center}}
\vspace{-0.6cm}
\caption{Energy balance in the simulation box, extracted from the PIC simulations for the target T2 and the three types of the laser pulses. The total energy, the electromagnetic energy and the particle energy are calculated in the simulation box and are all the energies per unit length. These energies are scaled to the total energy of the laser pulses.}\label{energy_bal}
\end{figure}

The efficiency of the laser energy conversion into the quasi-static magnetic field may be analyzed by the energy balance, shown in Fig. \ref{energy_bal} for the T2 target. The total energy deposited by the incident laser pulse in the simulation box grows approximately linearly during the laser pulse time, and it gradually decreases after the pulse ends, because the most energetic particles and the radiation leave the simulation box. Note, that the incident laser pulse energy is not shown in this figure. This explains fast oscillations of the total and electromagnetic energy in the case of a linear polarization. The absorbed laser energy is transferred first to electrons. Then, at the time scale of a hundred of femtoseconds, a part of this energy is transferred to ions due to the charge separation fields. The amount of the absorbed energy depends on the target length in $z$-direction, and can not be reproduced in the presented 2D3V simulations. Instead, we compare the energy of electrons with the total energy in the simulation box. According to Fig.\ref{energy_bal}, electrons and ions have $\sim15$\% and $\sim 60-70$\% of the maximum total energy in the simulation box for the shorter and longer pulses correspondingly. The electromagnetic energy during the simulation increases steadily with time, which is explained by the quasi-static magnetic field input. The residual part of the electromagnetic energy remaining in the simulation box after the laser pulse has gone then corresponds to the magnetic energy coupled to vortical plasma structures in the target hole. Estimation of the ratio between the magnetic field energy and the total energy in the simulation box (magnetization factor) gives the level of $10-20$\% depending on the interaction parameters. A more accurate estimate of the efficiency of magnetic field generation, defined as a ratio of the magnetic field energy to the total laser energy, requires a three-dimensional simulation and it depends on the target thickness in the laser propagation direction. For an optimal conditions, where a significant part of laser pulse energy is deposited in a target, is may reach a level of a few percent.

\section{Discussion}\label{sec3}

In this section, we separate the stage of generation of the magnetic filed, which is short and lasts as long as the laser pulse does, and the relaxation stage, where the magnetized structure slowly evolves. The first stage is also limited by the time of the internal plasma expansion. Because of this limitation, in our reduced approach, only short pulses have sense. For the case of our ``long'' pulse of 266 fs (see Figs. \ref{13-e__14-a__Bz}, \ref{13-e__edens_rj}, \ref{13-f_14-b__Bz}), the observed plasma expansion is still small.  

\subsection{Generation of the magnetic field}\label{sec3a}

Propagating inside the hollow of a chiral target, laser pulse heats the internal surface of the target. Electrons are gaining energy from a high-intensity laser pulse and leave the surface. As a secondary process, discharge currents of thermal electrons are excited, which tend to compensate the lack of electric charge on the surfaces of claws. Geometry of the claws defines the direction of the currents and in the case of a violated rotational symmetry leads to generation of long-living magnetized plasma structures in the target hole. 

The direction of the current of hot electrons is defined by the laser pulse parameters and it has a broad angular distribution. Later, when they are strongly deviated by the generated magnetic field, a self-consistent structure is formed. Opposite to the current of hot electrons, discharging currents are formed at both surfaces of the claws, and directed toward the central part of the void. In Figs. \ref{14__edens_rj} and \ref{13-e__14-a__Bz} at the early time moments, $200\tau_0$ and $75\tau_0$ respectively, the increase of electron density is seen around the sharp edge of the claws, closer to the center. These electrons are accelerated and then deviated by the self-generated magnetic field, as it is seen in Figs. \ref{14__edens_rj} and \ref{13-e__14-a__Bz} at later time moments of  $500\tau_0$ and $300\tau_0$ respectively. The discharge currents have the same direction at both edges of the claws. It can be deduced from Fig. \ref{13-d_14__Bz} at the time moment $200\tau_0$ and Fig. \ref{13-e__14-a__Bz} at the time moment $75\tau_0$, where the values of the magnetic field, generated near the claws, have the opposite directions at the opposite sides of the claws (compare regions $\alpha$ and $\beta$ in Fig. \ref{14-b__edens_Bz}). The spatial scale of the discharge currents is defined by the skin depth at the target claws, which is much smaller than their thickness. Because of a very sharp geometry of the discharge currents compared to the current of the hot electrons, the former ones are responsible for the generation of the magnetic field. This situation strongly differs from that considered in Ref. \cite{korneev-pre15}, where hot electrons were more collimated.  


%
%


We introduce the parameter of chirality $\chi_{tgt}$, defined by a target geometry and characterizing the efficiency of magnetic field generation by the surface currents. These currents are directed along the surface of claws and thus are defined by the target geometry. The magnetic field in the center $B_z(0)$ of a solenoid-type current distribution is defined by the Biot-Savart law  (we use CGS units below)
\beq
B_z(0)=\frac{2}{c}\int\limits_{\zeta} \dfrac{\mathbf J ({\zeta})\times \mathbf r({\zeta})}{r^2({\zeta})} ~d {\zeta},
\label{biot-savart}
\eeq
where $\zeta$ is a contour of claws, $\mathbf J ({\zeta}) =z^{-1} \int \mathbf j ({\mathbf r}) dr dz$ is the current on the claw contour per $z-$unit length, $\mathbf j ({\mathbf r})\sim\delta(\mathbf{r}-\vec{\zeta})$ is the claw surface current, and  ${r({\zeta})}$ is the distance to a surface point. 
The quantitative calculation of surface currents is a complex problem, which deals with the multiple parameters of laser-target interaction. To classify targets by their geometrical properties, we assume a constant absolute value of $\mathbf J ({\zeta})$, and normalize $B_z(0)$ in Eq. (\ref{biot-savart}) to the magnetic field created by the same current in the center of an ideal solenoid. Then the expression for the chirality parameter reads:
\beq
\chi_{tgt}=\dfrac{B_z(0)}{{4\pi J}/{c}}=\dfrac{1}{2\pi}\int\limits_{\zeta} \dfrac{\vec {\zeta}\times \mathbf r({\zeta})}{r^2({\zeta})} d {\zeta},
\label{def_eta_tgt}
\eeq
where $\vec \zeta$ is a unitary vector along the surface, directed as the current $J({\zeta})$.

Applying this definition to the targets T1 and T2 one finds:
\beq
\chi_{tgt}=\dfrac{1}{2\pi}\sum_{claws}\int_{ \theta _{min}}^{ \theta _{max}} d \theta \left({{\left( \dfrac{\Delta r}{r( \theta )  \theta _0} \right) ^2+1}}\right)^{-1/2}.
\label{calc_eta_tgt}
\eeq
After integration we obtain $\chi_{tgt}^{T1}=0.68$ for the target T1 and $\chi_{tgt}^{T2}=-1.34$ for the target T2. As expected, the sign of $\chi_{tgt}$ for these two targets is opposite, so that the magnetic field in the target center has the opposite direction. Also, from the calculated values of $\chi_{tgt}^{T1}$ and $\chi_{tgt}^{T2}$ it follows that the geometry of the target T2 is more suitable for the magnetic field generation. Indeed, as one can see in Figs. \ref{13-d_14__Bz},  \ref{13-e__14-a__Bz} and \ref{13-f_14-b__Bz}, in the target T2 magnetic field structures are more pronounced.

{
Scaling for the characteristic magnetic field $B_z(0)$ and the current $J$ with the laser intensity may be derived from the energy and charge conservation considerations. The currents of cold electrons on the claw surfaces are exited due to neutralization of a positive charge created under the action of a laser pulse, when the heated electrons are ejected almost isotropically from the interaction region. Then the value of surface currents can be estimated by the rate of hot electron ejection. Let the laser energy transmitted to hot electrons per unit time be $\eta_{e}P_{las}$, where $\eta_{e}$ is the transmission coefficient, $P_{las}$ is the total laser power. Assuming that it is distributed among the electrons with the mean energy $\epsilon_e$, the net discharge surface current per unit length is $J^*\sim e\eta_{e}P_{las}/\epsilon_e z_0$, where $z_0$ is the laser propagation length along $z-$axis in the target hole. 
This current, according to Eq. (\ref{biot-savart}), produces the magnetic field in the center of a target with the chirality $\chi_{tgt}$ 
 \beq
B_z^* \sim \chi_{tgt} \frac{4 \pi J^*}{c}\sim  \chi_{tgt} e\eta_{e} \frac{4\pi}{c z_0} \frac{P_{las}}{\epsilon_e}.
\label{B*}
\eeq
Using the ponderomotive scaling for electron energies for $a_0\gtrsim 1$ 
\beq
\epsilon_e\approx m_e c^2 \left(\sqrt{1+a_0^2}-1\right),
\label{energy_scaling}
\eeq
and for the parameters in Fig. \ref{13-d_14__Bz} we estimate the characteristic field as $B_z^*\sim  1.5\times 10^{8}\chi_{tgt}\eta_{e}/z_0$ Gauss, where $z_0$ is in cm. With the calculated chirality and the values of the magnetic fields, observed in the simulations, the coefficient $\eta_e/z_0$ appears to be $\sim 1 ~\mathrm{cm}^{-1}$. Assuming $\eta_e \sim5\%$ we obtain $z_0 \sim500~\mu$m. Such a relatively large value of $z_0$ supports the applicability of our 2D approach.
}

\subsection{Magnetized plasma structure in a target hole}\label{sec3b}

Let us consider the magnetized structures at the later times, when the pulse is gone. In Fig. \ref{energy_bal} this stage corresponds to times $t\gtrsim 0.5$ ps for 266 fs pulses, and $t\gtrsim 0.05$ ps for the 27 fs pulse. 

\begin{figure}
  \begin{tikzpicture}[x=0.51\linewidth,y=0.45\linewidth]
    \path
    (0,0) node{\includegraphics[width=0.5\linewidth]{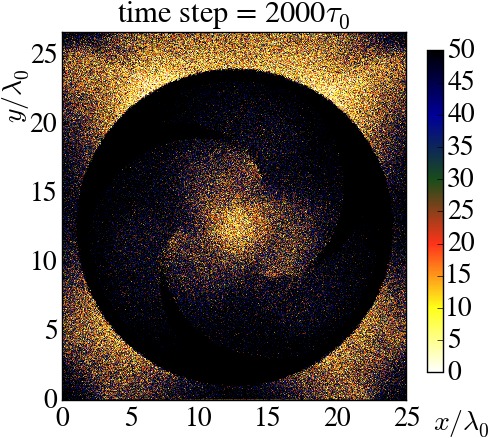}}

    (0.1,0.2) node { \textcolor{white}{\large{$n_e$}}}

    (1,0) node{\includegraphics[width=0.5\linewidth]{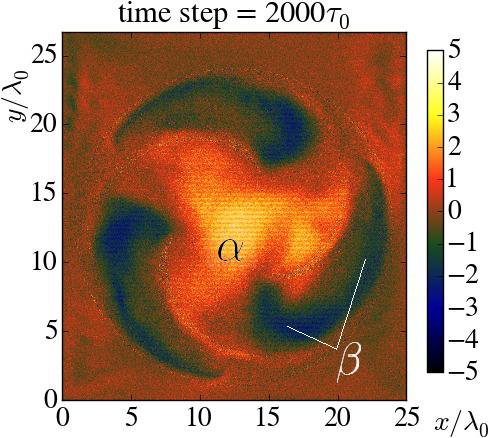}}

    (1.16,0.35) node { \textcolor{white}{\large{$B_z$}}}
;
  \end{tikzpicture}\caption{Electron density (left) and magnetic field (right) in the target T2 at the late time moment $2000\,\tau_0$, for a circularly polarized pump laser pulse with an intensity $10^{20}$W/cm$^2$ and duration 266 fs.} \label{14-b__edens_Bz}

\end{figure}

For the case of a 266 ps circularly polarized laser pulse, interacting with the target T2 at the late time moment $t=2000\tau_0\approx 5.3$ ps the electron density and the magnetic field distributions are shown in Fig. \ref{14-b__edens_Bz}. They correspond to a quasi-stationary magnetized structure. The profiles averaged over the polar angle for the $z-$component of the magnetic field $B_z(r)$, the electron density $ n_e(r)$, the $\theta-$component of the current density $j_\theta$, the charge density $Zn_i(r)-n_e(r)$, and the radial component electric field $E_r(r)$ are shown in Fig. \ref{profiles_2020}. The charge density and the electric field are relatively low compared to $n_e(r)$ and $B_z(r)$. Figure shows the magnetic field structure very similar to a $\theta-$pinch, which appears to be almost neutral with small radial electric field fluctuations. This special feature is related to non-relativistic velocities of electrons. The average positive current, coupled to the magnetic field, is generated due to the non-zero chirality of the target.

For the central magnetic structure formed inside a chiral target at late times, like the one shown in Fig. \ref{13-e__14-a__Bz} for the T2 target  at the time of $800\,\tau_0$, the plasma inside the hole is relatively cold and can be described by the two-fluid hydrodynamic equations:
\begin{subequations}
\begin{gather}
\partial_t n_{e,i} + \nabla ( n_{e,i} {\bf v}_{e,i} ) = 0, \\
\partial_t {\bf p}_{e,i}+({\bf v}_{e,i}\nabla) {\bf p}_{e,i} = q_{e,i}{\bf E} + \frac{q_{e,i}}{c}{\bf v}\times {\bf B},
\end{gather}  \label{generalHydro}
\end{subequations}
where ${\bf p}_{e,i}=\gamma m_{e,i} {\bf v}_{e,i}$ is the particle momentum, $\gamma=1/\sqrt{1-v^2/c^2}$ is the relativistic factor and $q_{e,i}=- e, Ze$ are the electron and ion charges. The electric and magnetic fields are described by the Maxwell's equations  with the current ${\bf j}=-e n_e {\bf v}_e + e Z n_i {\bf v}_i$ and the charge density $\rho = e Z n_i - e n_e $.
 
Let us consider stationary axially symmetric solutions corresponding to the electron vortex with immobile ions, the radial electric field $E_r(r)$, axial magnetic field $B_z(r)$ and the azimuthal electron velocity $v_\theta(r)$. Then the stationary equations read:
\begin{subequations}
\begin{gather}
 m_e \gamma v_\theta^2/r = e E_r +e v_\theta B_z/c , \label{1_charged_theta}\\
d_r B_z = 4\pi e n_e v_\theta/c, \label{2_charged_theta} \\
r^{-1}d_r (E_r r) = 4\pi e (Zn_i-n_e). \label{3_charged_theta}
\end{gather}
\label{charged_theta}
\end{subequations} 

These equations describe a collisionless $\theta-$pinch. The common nature between charged and neutral (collisional) $\theta-$pinch is plasma equilibrium at the given $r$, see Eq. \eqref{1_charged_theta}. In a conventional $\theta-$pinch the expanding force is the plasma pressure, while in our situation this role is played by the electric field between the electrons and immobile ions. As in the conventional $\theta$-pinch, there is a freedom to chose two functions due to the fact that the number of equations in (\ref{charged_theta}) is less than the number of variables. A particular solution depends on the scenario of electron beam production, it may be found as an initial-value time-dependent problem.    


We consider here model solutions by using the profiles, extracted from the PIC simulations for late times when the magnetized plasma structure has zero total charge
$Q_{tot}=0$, where
\beq
Q_{tot}=2\pi e\int\limits_0^\infty rdr(n_i-n_e).
\label{Q_tot}
\eeq
We assume then a profile of the magnetic field, similar to that in Fig. {\ref{profiles_2020}}
\begin{subequations}
\beq
B_z(r)=B_0\exp\left[ -r^2/r_0^2 \right],
\eeq
and the velocity profile, which evolves on the same spatial scale
\beq
v(r)=-v_0\frac{r}{r_0}\exp\left[ -r/r_0\right].
\eeq
Then, according to Eq. (\ref{2_charged_theta}) the electron density is
\beq
n(r)=\frac{2B_0}{v_0 r_0}\exp\left[ r/r_0-r^2/r_0^2 \right],
\label{n_neutral_theta}
\eeq
the electric field and the ion density are then defined from Eqs. (\ref{1_charged_theta}) and (\ref{3_charged_theta}) (the expressions are cumbersome to write them explicitly). The condition $Q_{tot}=0$, is fulfilled as long as $E_r(r)r\large\vert_{r=\infty}=0$.


\begin{figure}
{\begin{center}\includegraphics[width=9cm]{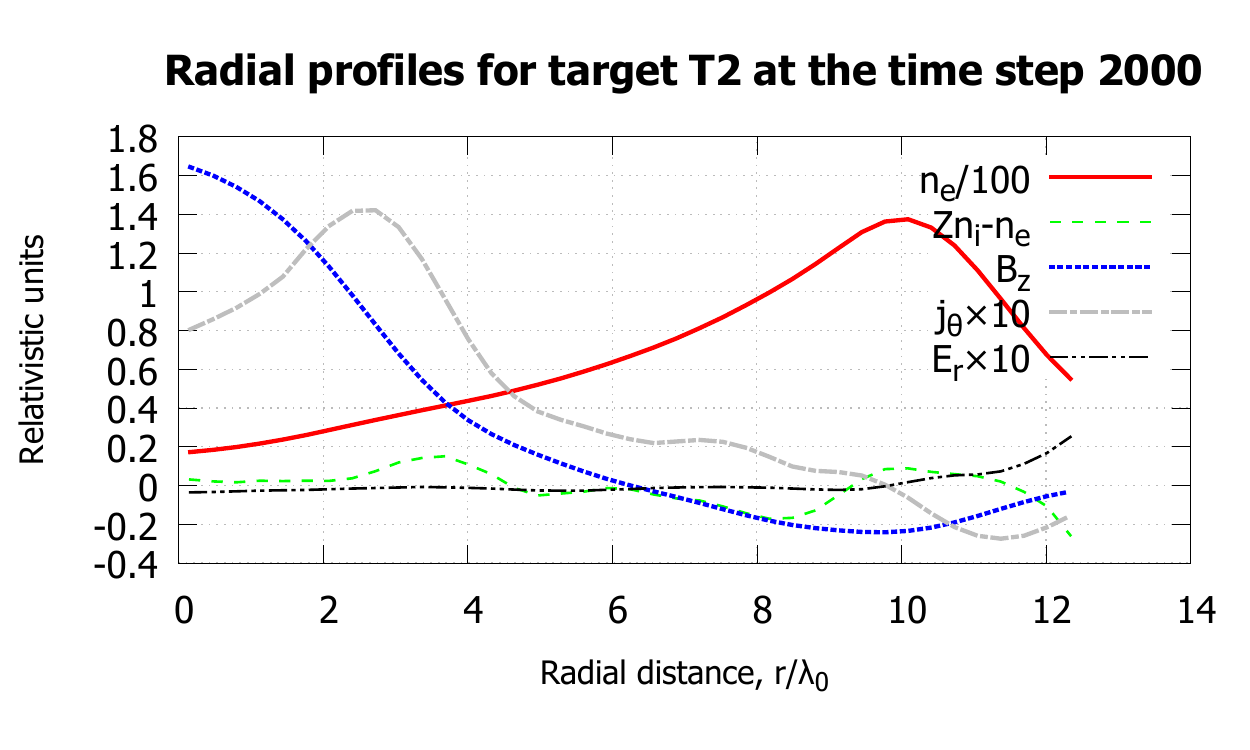}\end{center}}
\caption{Radial profiles for electron density $n(r)$, electric charge $Zn_i(r)-n_e(r)$, magnetic field $B_z$, tangential current component $j_\theta$ and radial component of electric field $E_r(r)$ for target T2 averaged over 100 fs around the time step $2000\tau_0$ ($t\approx 5.3$ ps) corresponding to Fig. \ref{14-b__edens_Bz}. 
}
\label{profiles_2020}
\end{figure}

%
%

\label{ex_neutral_theta}
\end{subequations}

\begin{figure}{\begin{center}\includegraphics[width=8cm,height=6cm]{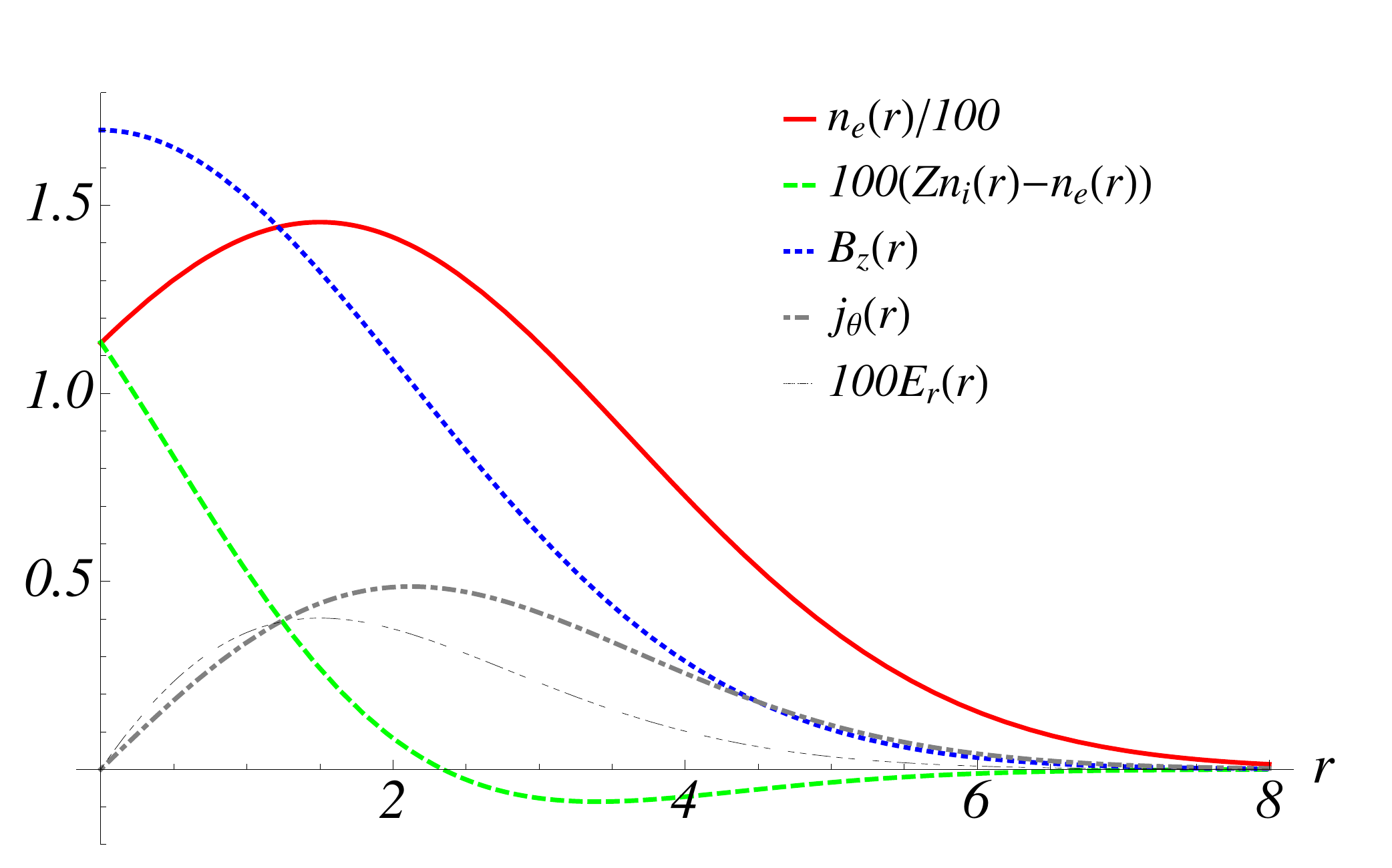}\end{center}}
\caption{Profiles for the fields, density, current and velocity, calculated from a system of equations (\ref{ex_neutral_theta}). The parameters are $B_0=1.7$ relativistic units, $r_0=3\lambda_0$, $v_0=0.01c$. Because of the very small electric field and the charge density, these two values are 100 times magnified, and because of the very high numbers for electron density, its value is reduced 100 times in the figure.}
\label{X-Profiles_TH0_1_X}
\end{figure}

The parameters in Eq. (\ref{ex_neutral_theta}), $B_0$, $r_0$ and $v_0$ are obtained from the profiles, shown in Fig.\ref{profiles_2020}. The characteristic scale length $r_0$ is of the order of 3 $\lambda_0$, and the amplitude $B_0\approx1.7$ in the relativistic units. The adjusted profiles are shown in Fig. \ref{X-Profiles_TH0_1_X}, they reproduce qualitatively the general features of the numerical results from Fig. \ref{profiles_2020}.

\begin{figure}{\begin{center}\includegraphics[width=9cm,height=6cm]{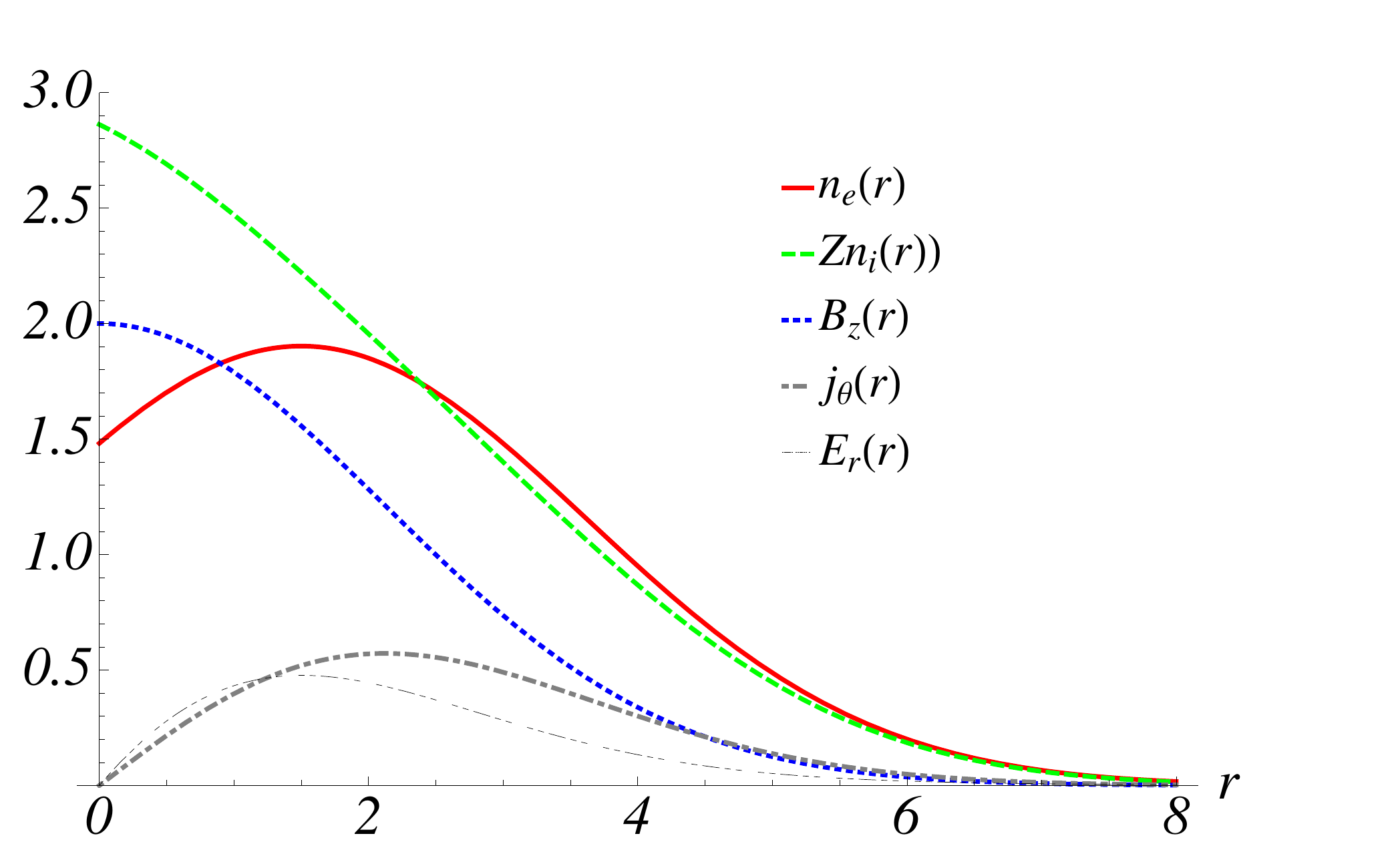}\end{center}}
\caption{Profiles for the fields, density, current and velocity, calculated from the system of equations (\ref{ex_neutral_theta}). The parameters are $B_0=2.0$ relativistic units, $r_0=3\lambda_0$, $v_0=0.9c$ to present a relativistic example of a $\theta-$pinch.}
\label{X-Profiles_TH0_2_X}
\end{figure}

The model presents in general a relativistic magnetized structure. To reach relativistic regime, the charge separation should be sufficient to produce a high electric field, which can bind relativistic electrons. For this situation, the profiles are shown in Fig. \ref{X-Profiles_TH0_2_X}, where we use another set of parameters $B_0=2$ relativistic units, $r_0=3\lambda_0$ and $v_0=0.9c$. As it follows from Fig. \ref{X-Profiles_TH0_2_X}, for the higher electron velocity the electron density is smaller. This is explained by Eq.(\ref{2_charged_theta}), which binds the current and the magnetic field. As a result of a higher velocity the centrifugal electrical force and the charge separation becomes greater.

\section{Conclusions and perspectives}\label{sec4}

We presented a scheme  for producing of strong quasi-stationary magnetic field structures in interaction of intense laser pulses with chiral targets. 
The proposition has some common features with the previous studies dealing with the laser generation of quasi-static magnetic field due to a special geometry of irradiated targets \cite{korneev-pre15, bigongiari-pop11}. Here, we make a step  in this direction toward more complex three-dimensional target geometries, which possess high symmetry and are more suitable for particle acceleration and astrophysical applications.  
\\ 
Magnetic fields in plasma are widely studied in the context of particle acceleration mechanisms \cite{jha-phrstab12, arefiev-jpp15}.
The magnetic field structure produced in the laser pulse interaction with a chiral target may be used for the guiding of a charged particle  beam. It would also be possible to accelerate charged particles directly from the magnetized plasma using a secondary laser pulse. 
As we showed, by choosing the 3D target structure one may control a spatial distribution of a magnetic field and use it for obtain certain conditions for particle or plasma beam production. 

In the considered approach, the plasma density may be varied by adjusting the size and material of a chiral target and a delay time between the laser pulses. In the case of low density magnetized structures, electrons could be accelerated using laser wakefield acceleration in the bubble regime \cite{Pukhov-apb02} and ions could be accelerated through collisionless shocks \cite{Fiuza-prl12, dHumieres-ppcf13}. In the case of higher density plasmas it would be possible to accelerate ions through charge-separation fields due to hot electrons exiting the target, like in the case of Target Normal Sheath Acceleration (TNSA) \cite{Wilks-pop01}, or directly through the laser radiation pressure, like in the case of Radiation Pressure Acceleration (RPA) \cite{Macchi-prl05, Robinson-njp08}. In all cases, the fact that the plasma in which the particles are accelerated is magnetized can lead to the generation of higher quality electron beams \cite{Vieira-prl11} than in the case of unmagnetized plasmas and to more energetic particle beams in the case of ion acceleration through collisionless shocks or TNSA as a better confinement of the hot electrons is expected in magnetized plasmas leading to higher accelerating fields. Laser ion acceleration through TNSA or RPA in such magnetized plasmas would also offer new possibilities to study magnetized plasma collisions in the laboratory, which are of great interest for laboratory astrophysics.


\section*{ACKNOWLEDGMENTS}
The work is in part supported by the French National funding agency ANR within the project SILAMPA, and it was granted access to the HPC resources of CINES under allocations 2015-056129 and 2016-056129 made by GENCI (Grand Equipement National de Calcul Intensif). A part of the numerical calculations was presented at Joint Supercomputer Center of the Russian Academy of Sciences and resources of NRNU MEPhI high-performance computing center. This work was supported by the Russian Presidential Grants for the Support of the Leading Scientific Schools 
and Russian Foundation for Basic Research (16-52-50019 \yaf).

%


\begin{thebibliography}{24}%
\makeatletter
\providecommand \@ifxundefined [1]{%
 \@ifx{#1\undefined}
}%
\providecommand \@ifnum [1]{%
 \ifnum #1\expandafter \@firstoftwo
 \else \expandafter \@secondoftwo
 \fi
}%
\providecommand \@ifx [1]{%
 \ifx #1\expandafter \@firstoftwo
 \else \expandafter \@secondoftwo
 \fi
}%
\providecommand \natexlab [1]{#1}%
\providecommand \enquote  [1]{``#1''}%
\providecommand \bibnamefont  [1]{#1}%
\providecommand \bibfnamefont [1]{#1}%
\providecommand \citenamefont [1]{#1}%
\providecommand \href@noop [0]{\@secondoftwo}%
\providecommand \href [0]{\begingroup \@sanitize@url \@href}%
\providecommand \@href[1]{\@@startlink{#1}\@@href}%
\providecommand \@@href[1]{\endgroup#1\@@endlink}%
\providecommand \@sanitize@url [0]{\catcode `\\12\catcode `\$12\catcode
  `\&12\catcode `\#12\catcode `\^12\catcode `\_12\catcode `\%12\relax}%
\providecommand \@@startlink[1]{}%
\providecommand \@@endlink[0]{}%
\providecommand \url  [0]{\begingroup\@sanitize@url \@url }%
\providecommand \@url [1]{\endgroup\@href {#1}{\urlprefix }}%
\providecommand \urlprefix  [0]{URL }%
\providecommand \Eprint [0]{\href }%
\providecommand \doibase [0]{http://dx.doi.org/}%
\providecommand \selectlanguage [0]{\@gobble}%
\providecommand \bibinfo  [0]{\@secondoftwo}%
\providecommand \bibfield  [0]{\@secondoftwo}%
\providecommand \translation [1]{[#1]}%
\providecommand \BibitemOpen [0]{}%
\providecommand \bibitemStop [0]{}%
\providecommand \bibitemNoStop [0]{.\EOS\space}%
\providecommand \EOS [0]{\spacefactor3000\relax}%
\providecommand \BibitemShut  [1]{\csname bibitem#1\endcsname}%
\let\auto@bib@innerbib\@empty
\bibitem [{\citenamefont {Esarey}\ \emph {et~al.}(2009)\citenamefont {Esarey},
  \citenamefont {Schroeder},\ and\ \citenamefont
  {Leemans}}]{esarey-revmodphys09}%
  \BibitemOpen
  \bibfield  {author} {\bibinfo {author} {\bibfnamefont {E.}~\bibnamefont
  {Esarey}}, \bibinfo {author} {\bibfnamefont {C.~B.}\ \bibnamefont
  {Schroeder}}, \ and\ \bibinfo {author} {\bibfnamefont {W.~P.}\ \bibnamefont
  {Leemans}},\ }\href {\doibase 10.1103/RevModPhys.81.1229} {\bibfield
  {journal} {\bibinfo  {journal} {Rev. Mod. Phys.}\ }\textbf {\bibinfo {volume}
  {81}},\ \bibinfo {pages} {1229} (\bibinfo {year} {2009})}\BibitemShut
  {NoStop}%
\bibitem [{\citenamefont {Macchi}\ \emph {et~al.}(2013)\citenamefont {Macchi},
  \citenamefont {Borghesi},\ and\ \citenamefont
  {Passoni}}]{macchi-revmodphys13}%
  \BibitemOpen
  \bibfield  {author} {\bibinfo {author} {\bibfnamefont {A.}~\bibnamefont
  {Macchi}}, \bibinfo {author} {\bibfnamefont {M.}~\bibnamefont {Borghesi}}, \
  and\ \bibinfo {author} {\bibfnamefont {M.}~\bibnamefont {Passoni}},\ }\href
  {\doibase 10.1103/RevModPhys.85.751} {\bibfield  {journal} {\bibinfo
  {journal} {Rev. Mod. Phys.}\ }\textbf {\bibinfo {volume} {85}},\ \bibinfo
  {pages} {751} (\bibinfo {year} {2013})}\BibitemShut {NoStop}%
\bibitem [{\citenamefont {Faure}\ \emph {et~al.}(2004)\citenamefont {Faure},
  \citenamefont {Glinec}, \citenamefont {Pukhov}, \citenamefont {Kiselev},
  \citenamefont {Gordienko}, \citenamefont {Lefebvre}, \citenamefont
  {Rousseau}, \citenamefont {Burgy},\ and\ \citenamefont
  {Malka}}]{pukhov-nature2004}%
  \BibitemOpen
  \bibfield  {author} {\bibinfo {author} {\bibfnamefont {J.}~\bibnamefont
  {Faure}}, \bibinfo {author} {\bibfnamefont {Y.}~\bibnamefont {Glinec}},
  \bibinfo {author} {\bibfnamefont {A.}~\bibnamefont {Pukhov}}, \bibinfo
  {author} {\bibfnamefont {S.}~\bibnamefont {Kiselev}}, \bibinfo {author}
  {\bibfnamefont {S.}~\bibnamefont {Gordienko}}, \bibinfo {author}
  {\bibfnamefont {E.}~\bibnamefont {Lefebvre}}, \bibinfo {author}
  {\bibfnamefont {J.-P.}\ \bibnamefont {Rousseau}}, \bibinfo {author}
  {\bibfnamefont {F.}~\bibnamefont {Burgy}}, \ and\ \bibinfo {author}
  {\bibfnamefont {V.}~\bibnamefont {Malka}},\ }\href@noop {} {\bibfield
  {journal} {\bibinfo  {journal} {Nature}\ }\textbf {\bibinfo {volume} {431}},\
  \bibinfo {pages} {541} (\bibinfo {year} {2004})}\BibitemShut {NoStop}%
\bibitem [{\citenamefont {Albertazzi}\ \emph {et~al.}(2014)\citenamefont
  {Albertazzi}, \citenamefont {Ciardi}, \citenamefont {Nakatsutsumi},
  \citenamefont {Vinci}, \citenamefont {B\'eard}, \citenamefont {Bonito},
  \citenamefont {Billette}, \citenamefont {Borghesi}, \citenamefont {Burkley},
  \citenamefont {Chen}, \citenamefont {Cowan}, \citenamefont
  {Herrmannsd\''{o}rfer}, \citenamefont {Higginson}, \citenamefont {Kroll},
  \citenamefont {Pikuz}, \citenamefont {Naughton}, \citenamefont {Romagnani},
  \citenamefont {Riconda}, \citenamefont {Revet}, \citenamefont {Riquier},
  \citenamefont {Schlenvoigt}, \citenamefont {Skobelev}, \citenamefont
  {Faenov}, \citenamefont {Soloviev}, \citenamefont {Huarte-Espinosa},
  \citenamefont {Frank}, \citenamefont {Portugall}, \citenamefont {P\'epin},\
  and\ \citenamefont {Fuchs}}]{albertazzi-science14}%
  \BibitemOpen
  \bibfield  {author} {\bibinfo {author} {\bibfnamefont {B.}~\bibnamefont
  {Albertazzi}}, \bibinfo {author} {\bibfnamefont {A.}~\bibnamefont {Ciardi}},
  \bibinfo {author} {\bibfnamefont {M.}~\bibnamefont {Nakatsutsumi}}, \bibinfo
  {author} {\bibfnamefont {T.}~\bibnamefont {Vinci}}, \bibinfo {author}
  {\bibfnamefont {J.}~\bibnamefont {B\'eard}}, \bibinfo {author} {\bibfnamefont
  {R.}~\bibnamefont {Bonito}}, \bibinfo {author} {\bibfnamefont
  {J.}~\bibnamefont {Billette}}, \bibinfo {author} {\bibfnamefont
  {M.}~\bibnamefont {Borghesi}}, \bibinfo {author} {\bibfnamefont
  {Z.}~\bibnamefont {Burkley}}, \bibinfo {author} {\bibfnamefont {S.~N.}\
  \bibnamefont {Chen}}, \bibinfo {author} {\bibfnamefont {T.~E.}\ \bibnamefont
  {Cowan}}, \bibinfo {author} {\bibfnamefont {T.}~\bibnamefont
  {Herrmannsd\''{o}rfer}}, \bibinfo {author} {\bibfnamefont {D.~P.}\
  \bibnamefont {Higginson}}, \bibinfo {author} {\bibfnamefont {F.}~\bibnamefont
  {Kroll}}, \bibinfo {author} {\bibfnamefont {S.~A.}\ \bibnamefont {Pikuz}},
  \bibinfo {author} {\bibfnamefont {K.}~\bibnamefont {Naughton}}, \bibinfo
  {author} {\bibfnamefont {L.}~\bibnamefont {Romagnani}}, \bibinfo {author}
  {\bibfnamefont {C.}~\bibnamefont {Riconda}}, \bibinfo {author} {\bibfnamefont
  {G.}~\bibnamefont {Revet}}, \bibinfo {author} {\bibfnamefont
  {R.}~\bibnamefont {Riquier}}, \bibinfo {author} {\bibfnamefont {H.-P.}\
  \bibnamefont {Schlenvoigt}}, \bibinfo {author} {\bibfnamefont {I.~Y.}\
  \bibnamefont {Skobelev}}, \bibinfo {author} {\bibfnamefont {A.}~\bibnamefont
  {Faenov}}, \bibinfo {author} {\bibfnamefont {A.}~\bibnamefont {Soloviev}},
  \bibinfo {author} {\bibfnamefont {M.}~\bibnamefont {Huarte-Espinosa}},
  \bibinfo {author} {\bibfnamefont {A.}~\bibnamefont {Frank}}, \bibinfo
  {author} {\bibfnamefont {O.}~\bibnamefont {Portugall}}, \bibinfo {author}
  {\bibfnamefont {H.}~\bibnamefont {P\'epin}}, \ and\ \bibinfo {author}
  {\bibfnamefont {J.}~\bibnamefont {Fuchs}},\ }\href {\doibase
  10.1126/science.1259694} {\bibfield  {journal} {\bibinfo  {journal}
  {Science}\ }\textbf {\bibinfo {volume} {346}},\ \bibinfo {pages} {325}
  (\bibinfo {year} {2014})},\ \Eprint
  {http://arxiv.org/abs/http://www.sciencemag.org/content/346/6207/325.full.pdf}
  {http://www.sciencemag.org/content/346/6207/325.full.pdf} \BibitemShut
  {NoStop}%
\bibitem [{\citenamefont {Santos}\ \emph {et~al.}(2015)\citenamefont {Santos},
  \citenamefont {Bailly-Grandvaux}, \citenamefont {Giuffrida}, \citenamefont
  {Forestier-Colleoni}, \citenamefont {Fujioka}, \citenamefont {Zhang},
  \citenamefont {Korneev}, \citenamefont {Bouillaud}, \citenamefont {Dorard},
  \citenamefont {Batani}, \citenamefont {Chevrot}, \citenamefont {Cross},
  \citenamefont {Crowston}, \citenamefont {Dubois}, \citenamefont {Gazave},
  \citenamefont {Gregori}, \citenamefont {d'Humi\'eres}, \citenamefont {Hulin},
  \citenamefont {Ishihara}, \citenamefont {Kojima}, \citenamefont {Loyez},
  \citenamefont {Marqu\`s}, \citenamefont {Morace}, \citenamefont {Nicola\"\i},
  \citenamefont {Peyrusse}, \citenamefont {Poy\'e}, \citenamefont {Raffestin},
  \citenamefont {Ribolzi}, \citenamefont {Roth}, \citenamefont {Schaumann},
  \citenamefont {Serres}, \citenamefont {Tikhonchuk}, \citenamefont {Vacar},\
  and\ \citenamefont {Woolsey}}]{santos-njp15}%
  \BibitemOpen
  \bibfield  {author} {\bibinfo {author} {\bibfnamefont {J.~J.}\ \bibnamefont
  {Santos}}, \bibinfo {author} {\bibfnamefont {M.}~\bibnamefont
  {Bailly-Grandvaux}}, \bibinfo {author} {\bibfnamefont {L.}~\bibnamefont
  {Giuffrida}}, \bibinfo {author} {\bibfnamefont {P.}~\bibnamefont
  {Forestier-Colleoni}}, \bibinfo {author} {\bibfnamefont {S.}~\bibnamefont
  {Fujioka}}, \bibinfo {author} {\bibfnamefont {Z.}~\bibnamefont {Zhang}},
  \bibinfo {author} {\bibfnamefont {P.}~\bibnamefont {Korneev}}, \bibinfo
  {author} {\bibfnamefont {R.}~\bibnamefont {Bouillaud}}, \bibinfo {author}
  {\bibfnamefont {S.}~\bibnamefont {Dorard}}, \bibinfo {author} {\bibfnamefont
  {D.}~\bibnamefont {Batani}}, \bibinfo {author} {\bibfnamefont
  {M.}~\bibnamefont {Chevrot}}, \bibinfo {author} {\bibfnamefont {J.~E.}\
  \bibnamefont {Cross}}, \bibinfo {author} {\bibfnamefont {R.}~\bibnamefont
  {Crowston}}, \bibinfo {author} {\bibfnamefont {J.-L.}\ \bibnamefont
  {Dubois}}, \bibinfo {author} {\bibfnamefont {J.}~\bibnamefont {Gazave}},
  \bibinfo {author} {\bibfnamefont {G.}~\bibnamefont {Gregori}}, \bibinfo
  {author} {\bibfnamefont {E.}~\bibnamefont {d'Humi\'eres}}, \bibinfo {author}
  {\bibfnamefont {S.}~\bibnamefont {Hulin}}, \bibinfo {author} {\bibfnamefont
  {K.}~\bibnamefont {Ishihara}}, \bibinfo {author} {\bibfnamefont
  {S.}~\bibnamefont {Kojima}}, \bibinfo {author} {\bibfnamefont
  {E.}~\bibnamefont {Loyez}}, \bibinfo {author} {\bibfnamefont {J.-R.}\
  \bibnamefont {Marqu\`s}}, \bibinfo {author} {\bibfnamefont {A.}~\bibnamefont
  {Morace}}, \bibinfo {author} {\bibfnamefont {P.}~\bibnamefont {Nicola\"\i}},
  \bibinfo {author} {\bibfnamefont {O.}~\bibnamefont {Peyrusse}}, \bibinfo
  {author} {\bibfnamefont {A.}~\bibnamefont {Poy\'e}}, \bibinfo {author}
  {\bibfnamefont {D.}~\bibnamefont {Raffestin}}, \bibinfo {author}
  {\bibfnamefont {J.}~\bibnamefont {Ribolzi}}, \bibinfo {author} {\bibfnamefont
  {M.}~\bibnamefont {Roth}}, \bibinfo {author} {\bibfnamefont {G.}~\bibnamefont
  {Schaumann}}, \bibinfo {author} {\bibfnamefont {F.}~\bibnamefont {Serres}},
  \bibinfo {author} {\bibfnamefont {V.~T.}\ \bibnamefont {Tikhonchuk}},
  \bibinfo {author} {\bibfnamefont {P.}~\bibnamefont {Vacar}}, \ and\ \bibinfo
  {author} {\bibfnamefont {N.}~\bibnamefont {Woolsey}},\ }\href
  {http://stacks.iop.org/1367-2630/17/i=8/a=083051} {\bibfield  {journal}
  {\bibinfo  {journal} {New Journal of Physics}\ }\textbf {\bibinfo {volume}
  {17}},\ \bibinfo {pages} {083051} (\bibinfo {year} {2015})}\BibitemShut
  {NoStop}%
\bibitem [{\citenamefont {Volpe}\ \emph {et~al.}(2014)\citenamefont {Volpe},
  \citenamefont {Feugeas}, \citenamefont {Nicolai}, \citenamefont {Santos},
  \citenamefont {Touati}, \citenamefont {Breil}, \citenamefont {Batani},\ and\
  \citenamefont {Tikhonchuk}}]{volpe-pre14}%
  \BibitemOpen
  \bibfield  {author} {\bibinfo {author} {\bibfnamefont {L.}~\bibnamefont
  {Volpe}}, \bibinfo {author} {\bibfnamefont {J.-L.}\ \bibnamefont {Feugeas}},
  \bibinfo {author} {\bibfnamefont {P.}~\bibnamefont {Nicolai}}, \bibinfo
  {author} {\bibfnamefont {J.~J.}\ \bibnamefont {Santos}}, \bibinfo {author}
  {\bibfnamefont {M.}~\bibnamefont {Touati}}, \bibinfo {author} {\bibfnamefont
  {J.}~\bibnamefont {Breil}}, \bibinfo {author} {\bibfnamefont
  {D.}~\bibnamefont {Batani}}, \ and\ \bibinfo {author} {\bibfnamefont
  {V.}~\bibnamefont {Tikhonchuk}},\ }\href {\doibase
  10.1103/PhysRevE.90.063108} {\bibfield  {journal} {\bibinfo  {journal} {Phys.
  Rev. E}\ }\textbf {\bibinfo {volume} {90}},\ \bibinfo {pages} {063108}
  (\bibinfo {year} {2014})}\BibitemShut {NoStop}%
\bibitem [{\citenamefont {Cai}\ \emph {et~al.}(2011)\citenamefont {Cai},
  \citenamefont {Zhu}, \citenamefont {Chen}, \citenamefont {Wu}, \citenamefont
  {He},\ and\ \citenamefont {Mima}}]{cai-pre11}%
  \BibitemOpen
  \bibfield  {author} {\bibinfo {author} {\bibfnamefont {H.-b.}\ \bibnamefont
  {Cai}}, \bibinfo {author} {\bibfnamefont {S.-p.}\ \bibnamefont {Zhu}},
  \bibinfo {author} {\bibfnamefont {M.}~\bibnamefont {Chen}}, \bibinfo {author}
  {\bibfnamefont {S.-z.}\ \bibnamefont {Wu}}, \bibinfo {author} {\bibfnamefont
  {X.~T.}\ \bibnamefont {He}}, \ and\ \bibinfo {author} {\bibfnamefont
  {K.}~\bibnamefont {Mima}},\ }\href {\doibase 10.1103/PhysRevE.83.036408}
  {\bibfield  {journal} {\bibinfo  {journal} {Phys. Rev. E}\ }\textbf {\bibinfo
  {volume} {83}},\ \bibinfo {pages} {036408} (\bibinfo {year}
  {2011})}\BibitemShut {NoStop}%
\bibitem [{\citenamefont {Robinson}\ and\ \citenamefont
  {Sherlock}(2007)}]{robinson-pop07}%
  \BibitemOpen
  \bibfield  {author} {\bibinfo {author} {\bibfnamefont {A.~P.~L.}\
  \bibnamefont {Robinson}}\ and\ \bibinfo {author} {\bibfnamefont
  {M.}~\bibnamefont {Sherlock}},\ }\href {\doibase
  http://dx.doi.org/10.1063/1.2768317} {\bibfield  {journal} {\bibinfo
  {journal} {Physics of Plasmas}\ }\textbf {\bibinfo {volume} {14}},\ \bibinfo
  {eid} {083105} (\bibinfo {year} {2007})}\BibitemShut {NoStop}%
\bibitem [{\citenamefont {Ramakrishna}\ \emph {et~al.}(2010)\citenamefont
  {Ramakrishna}, \citenamefont {Kar}, \citenamefont {Robinson}, \citenamefont
  {Adams}, \citenamefont {Markey}, \citenamefont {Quinn}, \citenamefont {Yuan},
  \citenamefont {McKenna}, \citenamefont {Lancaster}, \citenamefont {Green},
  \citenamefont {Scott}, \citenamefont {Norreys}, \citenamefont {Schreiber},\
  and\ \citenamefont {Zepf}}]{ramakrishna-prl10}%
  \BibitemOpen
  \bibfield  {author} {\bibinfo {author} {\bibfnamefont {B.}~\bibnamefont
  {Ramakrishna}}, \bibinfo {author} {\bibfnamefont {S.}~\bibnamefont {Kar}},
  \bibinfo {author} {\bibfnamefont {A.~P.~L.}\ \bibnamefont {Robinson}},
  \bibinfo {author} {\bibfnamefont {D.~J.}\ \bibnamefont {Adams}}, \bibinfo
  {author} {\bibfnamefont {K.}~\bibnamefont {Markey}}, \bibinfo {author}
  {\bibfnamefont {M.~N.}\ \bibnamefont {Quinn}}, \bibinfo {author}
  {\bibfnamefont {X.~H.}\ \bibnamefont {Yuan}}, \bibinfo {author}
  {\bibfnamefont {P.}~\bibnamefont {McKenna}}, \bibinfo {author} {\bibfnamefont
  {K.~L.}\ \bibnamefont {Lancaster}}, \bibinfo {author} {\bibfnamefont {J.~S.}\
  \bibnamefont {Green}}, \bibinfo {author} {\bibfnamefont {R.~H.~H.}\
  \bibnamefont {Scott}}, \bibinfo {author} {\bibfnamefont {P.~A.}\ \bibnamefont
  {Norreys}}, \bibinfo {author} {\bibfnamefont {J.}~\bibnamefont {Schreiber}},
  \ and\ \bibinfo {author} {\bibfnamefont {M.}~\bibnamefont {Zepf}},\ }\href
  {\doibase 10.1103/PhysRevLett.105.135001} {\bibfield  {journal} {\bibinfo
  {journal} {Phys. Rev. Lett.}\ }\textbf {\bibinfo {volume} {105}},\ \bibinfo
  {pages} {135001} (\bibinfo {year} {2010})}\BibitemShut {NoStop}%
\bibitem [{\citenamefont {Shi}\ \emph {et~al.}(2014)\citenamefont {Shi},
  \citenamefont {Shen}, \citenamefont {Zhang}, \citenamefont {Zhang},
  \citenamefont {Wang},\ and\ \citenamefont {Xu}}]{shi-prl14}%
  \BibitemOpen
  \bibfield  {author} {\bibinfo {author} {\bibfnamefont {Y.}~\bibnamefont
  {Shi}}, \bibinfo {author} {\bibfnamefont {B.}~\bibnamefont {Shen}}, \bibinfo
  {author} {\bibfnamefont {L.}~\bibnamefont {Zhang}}, \bibinfo {author}
  {\bibfnamefont {X.}~\bibnamefont {Zhang}}, \bibinfo {author} {\bibfnamefont
  {W.}~\bibnamefont {Wang}}, \ and\ \bibinfo {author} {\bibfnamefont
  {Z.}~\bibnamefont {Xu}},\ }\href {\doibase 10.1103/PhysRevLett.112.235001}
  {\bibfield  {journal} {\bibinfo  {journal} {Phys. Rev. Lett.}\ }\textbf
  {\bibinfo {volume} {112}},\ \bibinfo {pages} {235001} (\bibinfo {year}
  {2014})}\BibitemShut {NoStop}%
\bibitem [{\citenamefont {L{\`e}cz}\ \emph {et~al.}(2015)\citenamefont
  {L{\`e}cz}, \citenamefont {Andreev},\ and\ \citenamefont
  {Seryi}}]{lecz-lpb15}%
  \BibitemOpen
  \bibfield  {author} {\bibinfo {author} {\bibfnamefont {Z.}~\bibnamefont
  {L{\`e}cz}}, \bibinfo {author} {\bibfnamefont {A.}~\bibnamefont {Andreev}}, \
  and\ \bibinfo {author} {\bibfnamefont {A.}~\bibnamefont {Seryi}},\ }\href
  {\doibase 10.1017/S0263034615000853} {\bibfield  {journal} {\bibinfo
  {journal} {Laser and Particle Beams}\ }\textbf {\bibinfo {volume}
  {FirstView}},\ \bibinfo {pages} {1} (\bibinfo {year} {2015})}\BibitemShut
  {NoStop}%
\bibitem [{\citenamefont {Korneev}\ \emph {et~al.}(2015)\citenamefont
  {Korneev}, \citenamefont {d'Humi{\`e}res},\ and\ \citenamefont
  {Tikhonchuk}}]{korneev-pre15}%
  \BibitemOpen
  \bibfield  {author} {\bibinfo {author} {\bibfnamefont {P.}~\bibnamefont
  {Korneev}}, \bibinfo {author} {\bibfnamefont {E.}~\bibnamefont
  {d'Humi{\`e}res}}, \ and\ \bibinfo {author} {\bibfnamefont {V.}~\bibnamefont
  {Tikhonchuk}},\ }\href@noop {} {\bibfield  {journal} {\bibinfo  {journal}
  {Physical Review E}\ }\textbf {\bibinfo {volume} {91}},\ \bibinfo {pages}
  {043107} (\bibinfo {year} {2015})}\BibitemShut {NoStop}%
\bibitem [{\citenamefont {Sentoku}\ and\ \citenamefont
  {Kemp}(2008)}]{Sentoku-jcp08}%
  \BibitemOpen
  \bibfield  {author} {\bibinfo {author} {\bibfnamefont {Y.}~\bibnamefont
  {Sentoku}}\ and\ \bibinfo {author} {\bibfnamefont {A.~J.}\ \bibnamefont
  {Kemp}},\ }\href {\doibase 10.1016/j.jcp.2008.03.043} {\bibfield  {journal}
  {\bibinfo  {journal} {J. Comput. Phys.}\ }\textbf {\bibinfo {volume} {227}},\
  \bibinfo {pages} {6846} (\bibinfo {year} {2008})}\BibitemShut {NoStop}%
\bibitem [{\citenamefont {Takizuka}\ and\ \citenamefont
  {Abe}(1977)}]{Takizuka-jcp77}%
  \BibitemOpen
  \bibfield  {author} {\bibinfo {author} {\bibfnamefont {T.}~\bibnamefont
  {Takizuka}}\ and\ \bibinfo {author} {\bibfnamefont {H.}~\bibnamefont {Abe}},\
  }\href {\doibase http://dx.doi.org/10.1016/0021-9991(77)90099-7} {\bibfield
  {journal} {\bibinfo  {journal} {Journal of Computational Physics}\ }\textbf
  {\bibinfo {volume} {25}},\ \bibinfo {pages} {205 } (\bibinfo {year}
  {1977})}\BibitemShut {NoStop}%
\bibitem [{\citenamefont {Bigongiari}\ \emph {et~al.}(2011)\citenamefont
  {Bigongiari}, \citenamefont {Raynaud}, \citenamefont {Riconda}, \citenamefont
  {H{\`e}ron},\ and\ \citenamefont {Macchi}}]{bigongiari-pop11}%
  \BibitemOpen
  \bibfield  {author} {\bibinfo {author} {\bibfnamefont {A.}~\bibnamefont
  {Bigongiari}}, \bibinfo {author} {\bibfnamefont {M.}~\bibnamefont {Raynaud}},
  \bibinfo {author} {\bibfnamefont {C.}~\bibnamefont {Riconda}}, \bibinfo
  {author} {\bibfnamefont {A.}~\bibnamefont {H{\`e}ron}}, \ and\ \bibinfo
  {author} {\bibfnamefont {A.}~\bibnamefont {Macchi}},\ }\href {\doibase
  http://dx.doi.org/10.1063/1.3646520} {\bibfield  {journal} {\bibinfo
  {journal} {Physics of Plasmas}\ }\textbf {\bibinfo {volume} {18}},\ \bibinfo
  {eid} {102701} (\bibinfo {year} {2011}),\
  http://dx.doi.org/10.1063/1.3646520}\BibitemShut {NoStop}%
\bibitem [{\citenamefont {Jha}\ \emph {et~al.}(2012)\citenamefont {Jha},
  \citenamefont {Saroch}, \citenamefont {Mishra},\ and\ \citenamefont
  {Upadhyay}}]{jha-phrstab12}%
  \BibitemOpen
  \bibfield  {author} {\bibinfo {author} {\bibfnamefont {P.}~\bibnamefont
  {Jha}}, \bibinfo {author} {\bibfnamefont {A.}~\bibnamefont {Saroch}},
  \bibinfo {author} {\bibfnamefont {R.~K.}\ \bibnamefont {Mishra}}, \ and\
  \bibinfo {author} {\bibfnamefont {A.~K.}\ \bibnamefont {Upadhyay}},\ }\href
  {\doibase 10.1103/PhysRevSTAB.15.081301} {\bibfield  {journal} {\bibinfo
  {journal} {Phys. Rev. ST Accel. Beams}\ }\textbf {\bibinfo {volume} {15}},\
  \bibinfo {pages} {081301} (\bibinfo {year} {2012})}\BibitemShut {NoStop}%
\bibitem [{\citenamefont {Arefiev}\ \emph {et~al.}(2015)\citenamefont
  {Arefiev}, \citenamefont {Robinson},\ and\ \citenamefont
  {Khudik}}]{arefiev-jpp15}%
  \BibitemOpen
  \bibfield  {author} {\bibinfo {author} {\bibfnamefont {A.~V.}\ \bibnamefont
  {Arefiev}}, \bibinfo {author} {\bibfnamefont {A.~P.~L.}\ \bibnamefont
  {Robinson}}, \ and\ \bibinfo {author} {\bibfnamefont {V.~N.}\ \bibnamefont
  {Khudik}},\ }\href {\doibase 10.1017/S0022377815000434} {\bibfield  {journal}
  {\bibinfo  {journal} {Journal of Plasma Physics}\ }\textbf {\bibinfo {volume}
  {81}} (\bibinfo {year} {2015}),\ 10.1017/S0022377815000434}\BibitemShut
  {NoStop}%
\bibitem [{\citenamefont {Pukhov}\ and\ \citenamefont {Meyer-ter
  Vehn}(2002)}]{Pukhov-apb02}%
  \BibitemOpen
  \bibfield  {author} {\bibinfo {author} {\bibfnamefont {A.}~\bibnamefont
  {Pukhov}}\ and\ \bibinfo {author} {\bibfnamefont {J.}~\bibnamefont {Meyer-ter
  Vehn}},\ }\href@noop {} {\bibfield  {journal} {\bibinfo  {journal} {Applied
  Physics B}\ }\textbf {\bibinfo {volume} {74}},\ \bibinfo {pages} {355}
  (\bibinfo {year} {2002})}\BibitemShut {NoStop}%
\bibitem [{\citenamefont {Fiuza}\ \emph {et~al.}(2012)\citenamefont {Fiuza},
  \citenamefont {Fonseca}, \citenamefont {Tonge}, \citenamefont {Mori},\ and\
  \citenamefont {Silva}}]{Fiuza-prl12}%
  \BibitemOpen
  \bibfield  {author} {\bibinfo {author} {\bibfnamefont {F.}~\bibnamefont
  {Fiuza}}, \bibinfo {author} {\bibfnamefont {R.~A.}\ \bibnamefont {Fonseca}},
  \bibinfo {author} {\bibfnamefont {J.}~\bibnamefont {Tonge}}, \bibinfo
  {author} {\bibfnamefont {W.~B.}\ \bibnamefont {Mori}}, \ and\ \bibinfo
  {author} {\bibfnamefont {L.~O.}\ \bibnamefont {Silva}},\ }\href {\doibase
  10.1103/PhysRevLett.108.235004} {\bibfield  {journal} {\bibinfo  {journal}
  {Phys. Rev. Lett.}\ }\textbf {\bibinfo {volume} {108}},\ \bibinfo {pages}
  {235004} (\bibinfo {year} {2012})}\BibitemShut {NoStop}%
\bibitem [{\citenamefont {d'Humi{\`e}res}\ \emph {et~al.}(2013)\citenamefont
  {d'Humi{\`e}res}, \citenamefont {Antici}, \citenamefont {Glesser},
  \citenamefont {Boeker}, \citenamefont {Cardelli}, \citenamefont {Chen},
  \citenamefont {Feugeas}, \citenamefont {Filippi}, \citenamefont {Gauthier},
  \citenamefont {Levy}, \citenamefont {Nicola\"\i}, \citenamefont {P{\'e}pin},
  \citenamefont {Romagnani}, \citenamefont {Scisci{\`o}}, \citenamefont
  {Tikhonchuk}, \citenamefont {Willi}, \citenamefont {Kieffer},\ and\
  \citenamefont {Fuchs}}]{dHumieres-ppcf13}%
  \BibitemOpen
  \bibfield  {author} {\bibinfo {author} {\bibfnamefont {E.}~\bibnamefont
  {d'Humi{\`e}res}}, \bibinfo {author} {\bibfnamefont {P.}~\bibnamefont
  {Antici}}, \bibinfo {author} {\bibfnamefont {M.}~\bibnamefont {Glesser}},
  \bibinfo {author} {\bibfnamefont {J.}~\bibnamefont {Boeker}}, \bibinfo
  {author} {\bibfnamefont {F.}~\bibnamefont {Cardelli}}, \bibinfo {author}
  {\bibfnamefont {S.}~\bibnamefont {Chen}}, \bibinfo {author} {\bibfnamefont
  {J.~L.}\ \bibnamefont {Feugeas}}, \bibinfo {author} {\bibfnamefont
  {F.}~\bibnamefont {Filippi}}, \bibinfo {author} {\bibfnamefont
  {M.}~\bibnamefont {Gauthier}}, \bibinfo {author} {\bibfnamefont
  {A.}~\bibnamefont {Levy}}, \bibinfo {author} {\bibfnamefont {P.}~\bibnamefont
  {Nicola\"\i}}, \bibinfo {author} {\bibfnamefont {H.}~\bibnamefont
  {P{\'e}pin}}, \bibinfo {author} {\bibfnamefont {L.}~\bibnamefont
  {Romagnani}}, \bibinfo {author} {\bibfnamefont {M.}~\bibnamefont
  {Scisci{\`o}}}, \bibinfo {author} {\bibfnamefont {V.~T.}\ \bibnamefont
  {Tikhonchuk}}, \bibinfo {author} {\bibfnamefont {O.}~\bibnamefont {Willi}},
  \bibinfo {author} {\bibfnamefont {J.~C.}\ \bibnamefont {Kieffer}}, \ and\
  \bibinfo {author} {\bibfnamefont {J.}~\bibnamefont {Fuchs}},\ }\href
  {http://stacks.iop.org/0741-3335/55/i=12/a=124025} {\bibfield  {journal}
  {\bibinfo  {journal} {Plasma Physics and Controlled Fusion}\ }\textbf
  {\bibinfo {volume} {55}},\ \bibinfo {pages} {124025} (\bibinfo {year}
  {2013})}\BibitemShut {NoStop}%
\bibitem [{\citenamefont {Wilks}\ \emph {et~al.}(2001)\citenamefont {Wilks},
  \citenamefont {Langdon}, \citenamefont {Cowan}, \citenamefont {Roth},
  \citenamefont {Singh}, \citenamefont {Hatchett}, \citenamefont {Key},
  \citenamefont {Pennington}, \citenamefont {MacKinnon},\ and\ \citenamefont
  {Snavely}}]{Wilks-pop01}%
  \BibitemOpen
  \bibfield  {author} {\bibinfo {author} {\bibfnamefont {S.~C.}\ \bibnamefont
  {Wilks}}, \bibinfo {author} {\bibfnamefont {A.~B.}\ \bibnamefont {Langdon}},
  \bibinfo {author} {\bibfnamefont {T.~E.}\ \bibnamefont {Cowan}}, \bibinfo
  {author} {\bibfnamefont {M.}~\bibnamefont {Roth}}, \bibinfo {author}
  {\bibfnamefont {M.}~\bibnamefont {Singh}}, \bibinfo {author} {\bibfnamefont
  {S.}~\bibnamefont {Hatchett}}, \bibinfo {author} {\bibfnamefont {M.~H.}\
  \bibnamefont {Key}}, \bibinfo {author} {\bibfnamefont {D.}~\bibnamefont
  {Pennington}}, \bibinfo {author} {\bibfnamefont {A.}~\bibnamefont
  {MacKinnon}}, \ and\ \bibinfo {author} {\bibfnamefont {R.~A.}\ \bibnamefont
  {Snavely}},\ }\href {\doibase http://dx.doi.org/10.1063/1.1333697} {\bibfield
   {journal} {\bibinfo  {journal} {Physics of Plasmas (1994-present)}\ }\textbf
  {\bibinfo {volume} {8}},\ \bibinfo {pages} {542} (\bibinfo {year}
  {2001})}\BibitemShut {NoStop}%
\bibitem [{\citenamefont {Macchi}\ \emph {et~al.}(2005)\citenamefont {Macchi},
  \citenamefont {Cattani}, \citenamefont {Liseykina},\ and\ \citenamefont
  {Cornolti}}]{Macchi-prl05}%
  \BibitemOpen
  \bibfield  {author} {\bibinfo {author} {\bibfnamefont {A.}~\bibnamefont
  {Macchi}}, \bibinfo {author} {\bibfnamefont {F.}~\bibnamefont {Cattani}},
  \bibinfo {author} {\bibfnamefont {T.~V.}\ \bibnamefont {Liseykina}}, \ and\
  \bibinfo {author} {\bibfnamefont {F.}~\bibnamefont {Cornolti}},\ }\href
  {\doibase 10.1103/PhysRevLett.94.165003} {\bibfield  {journal} {\bibinfo
  {journal} {Phys. Rev. Lett.}\ }\textbf {\bibinfo {volume} {94}},\ \bibinfo
  {pages} {165003} (\bibinfo {year} {2005})}\BibitemShut {NoStop}%
\bibitem [{\citenamefont {Robinson}\ \emph {et~al.}(2008)\citenamefont
  {Robinson}, \citenamefont {Zepf}, \citenamefont {Kar}, \citenamefont
  {Evans},\ and\ \citenamefont {Bellei}}]{Robinson-njp08}%
  \BibitemOpen
  \bibfield  {author} {\bibinfo {author} {\bibfnamefont {A.~P.~L.}\
  \bibnamefont {Robinson}}, \bibinfo {author} {\bibfnamefont {M.}~\bibnamefont
  {Zepf}}, \bibinfo {author} {\bibfnamefont {S.}~\bibnamefont {Kar}}, \bibinfo
  {author} {\bibfnamefont {R.~G.}\ \bibnamefont {Evans}}, \ and\ \bibinfo
  {author} {\bibfnamefont {C.}~\bibnamefont {Bellei}},\ }\href
  {http://stacks.iop.org/1367-2630/10/i=1/a=013021} {\bibfield  {journal}
  {\bibinfo  {journal} {New Journal of Physics}\ }\textbf {\bibinfo {volume}
  {10}},\ \bibinfo {pages} {013021} (\bibinfo {year} {2008})}\BibitemShut
  {NoStop}%
\bibitem [{\citenamefont {Vieira}\ \emph {et~al.}(2011)\citenamefont {Vieira},
  \citenamefont {Martins}, \citenamefont {Pathak}, \citenamefont {Fonseca},
  \citenamefont {Mori},\ and\ \citenamefont {Silva}}]{Vieira-prl11}%
  \BibitemOpen
  \bibfield  {author} {\bibinfo {author} {\bibfnamefont {J.}~\bibnamefont
  {Vieira}}, \bibinfo {author} {\bibfnamefont {S.~F.}\ \bibnamefont {Martins}},
  \bibinfo {author} {\bibfnamefont {V.~B.}\ \bibnamefont {Pathak}}, \bibinfo
  {author} {\bibfnamefont {R.~A.}\ \bibnamefont {Fonseca}}, \bibinfo {author}
  {\bibfnamefont {W.~B.}\ \bibnamefont {Mori}}, \ and\ \bibinfo {author}
  {\bibfnamefont {L.~O.}\ \bibnamefont {Silva}},\ }\href {\doibase
  10.1103/PhysRevLett.106.225001} {\bibfield  {journal} {\bibinfo  {journal}
  {Phys. Rev. Lett.}\ }\textbf {\bibinfo {volume} {106}},\ \bibinfo {pages}
  {225001} (\bibinfo {year} {2011})}\BibitemShut {NoStop}%
\end{thebibliography}

%
%
%
%
%
%
%
%
%
%
\end{document}